\documentclass[useAMS,usenatbib,usegraphicx]{mn2e}
\usepackage{times}
\usepackage{amsmath}

\title[Bivariate Galaxy Luminosity Functions in the SDSS]{Bivariate Galaxy Luminosity Functions in the Sloan Digital Sky Survey}

\author[N. M. Ball et al.]{N. M. Ball,$^{1,2}$\thanks{E-mail: nball@astro.uiuc.edu} J. Loveday,$^{3}$ R. J. Brunner,$^{1,2}$ I.K. Baldry$^{4}$ and J. Brinkmann$^{5}$\\ $^{1}$Department of Astronomy, MC-221, University of Illinois, 1002 West Green Street, Urbana, IL 61801, USA \\ $^{2}$National Center for Supercomputing Applications, MC-476, University of Illinois, 605 East Springfield Avenue, Champaign, IL 61820, USA\\ $^{3}$Astronomy Centre, University of Sussex, Falmer, Brighton, BN1 9QJ, UK\\ $^{4}$Astrophysics Research Institute, Liverpool John Moores University, Twelve Quays House, Egerton Wharf, Birkenhead CH41 1LD, UK\\ $^{5}$Apache Point Observatory, PO Box 59, Sunspot, NM 88349, USA}

\newcommand{\mrm}{\mathrm}

\newcommand{\omegam}{\Omega_{\mrm{matter}}}
\newcommand{\omegal}{\Omega_{\mrm{\Lambda}}}
\newcommand{\mrlogh}{M_r - 5 {\mrm{log}} h}
\newcommand{\sbrapp}{\mu_{\mrm{app}}}

\newcommand{\ciinv}{CI_{\mrm{inv}}}
\newcommand{\sbr}{\mu_{\mrm{R_{50}}}}
\newcommand{\absrad}{R_{\mrm{abs}}}
\newcommand{\tann}{T_{\mrm{ANN}}}
\newcommand{\tjpg}{T_{\mrm{JPG}}}
\newcommand{\eclass}{\mrm{eClass}}
\newcommand{\mstellar}{M_{\mrm{stellar}}}
\newcommand{\logn}{\mrm{log}~n}

\newcommand{\kms}{~\mrm{km~s^{-1}}}
\newcommand{\kmsp}{~\mrm{km~s^{-1}~Mpc^{-1}}}
\newcommand{\sqdeg}{~\mrm{deg}^2}
\newcommand{\magarcsec}{~\mrm{mag~arcsec}^{-2}}
\newcommand{\mgn}{~\mrm{mag}} 
\newcommand{\msun}{~\mrm{M}_{\sun}}
\newcommand{\dens}{~h_{70}^{2}~{\mrm{Mpc}}^{-2}}
\newcommand{\kpc}{~h^{-1}~{\mrm{kpc}}}
\newcommand{\cubicmpc}{~h^{-3}~{\mrm{Mpc}}^3}

\newcommand{\aap}{A\&A}
\newcommand{\aj}{AJ}

\newcommand{\apj}{ApJ}
\newcommand{\apjl}{ApJL}
\newcommand{\apjs}{ApJS}
\newcommand{\araa}{ARA\&A}
\newcommand{\mnras}{MNRAS}

\newcommand{\pasa}{PASA}

\newcommand{\pasp}{PASP}

\def\spose#1{\hbox to 0pt{#1\hss}}
\def\simlt{\mathrel{\spose{\lower 3pt\hbox{$\mathchar"218$}}
     \raise 2.0pt\hbox{$\mathchar"13C$}}}
\def\simgt{\mathrel{\spose{\lower 3pt\hbox{$\mathchar"218$}}
     \raise 2.0pt\hbox{$\mathchar"13E$}}}

\begin{document}

\date{Accepted xxxx Received xxxx}

\pagerange{\pageref{firstpage}--\pageref{lastpage}} \pubyear{2006}

\maketitle

\label{firstpage}

\begin{abstract}
Bivariate luminosity functions (LFs) are computed for galaxies in the New York Value-Added Galaxy Catalogue, based on the Sloan Digital Sky Survey Data Release 4. The galaxy properties investigated are the morphological type, inverse concentration index, S\'ersic index, absolute effective surface brightness, reference frame colours, absolute radius, eClass spectral type, stellar mass and galaxy environment. The morphological sample is flux-limited to galaxies with $r<15.9$ and consists of 37,047 classifications to an RMS accuracy of $\pm$ half a class in the sequence E, S0, Sa, Sb, Sc, Sd, Im. These were assigned by an artificial neural network, based on a training set of 645 eyeball classifications. The other samples use $r<17.77$ with a median redshift of $z \sim 0.08$, and a limiting redshift of $z < 0.15$ to minimize the effects of evolution. Other cuts, for example in axis ratio, are made to minimize biases. A wealth of detail is seen, with clear variations between the LFs according to absolute magnitude and the second parameter. They are consistent with an early type, bright, concentrated, red population and a late type, faint, less concentrated, blue, star forming population. This bimodality suggests two major underlying physical processes, which in agreement with previous authors we hypothesize to be merger and accretion, associated with the properties of bulges and discs respectively. The bivariate luminosity-surface brightness distribution is fit with the Cho{\l}oniewski function (a Schechter function in absolute magnitude and Gaussian in surface brightness). The fit is found to be poor, as might be expected if there are two underlying processes.
\end{abstract}


\section{Introduction} \label{sec: intro}

The galaxy luminosity function (LF) is well-known as a fundamental measurement of the properties of galaxies. It constrains theories of their formation and evolution, such as biased galaxy formation \citep[e.g.][]{benson:bias} and is needed for many other measurements. Examples are the luminosity density in the universe \citep[e.g.][]{cross:ldens}, the selection function in magnitude-limited galaxy surveys \citep[e.g.][]{norberg:lf}, the distances to various types of object, the number of absorbing objects subdivided by redshift and deprojecting the angular correlation function from two dimensions to three \citep[e.g.][]{limber:deprojection, binggeli:lf}.

The general or universal luminosity function, i.e. that for all galaxies, is defined as the comoving number density of galaxies from luminosity $L$ to $L+dL$ \begin{equation} dN = \phi(L)~dL~dV . \end{equation}

Measurements of the universal form began with \citet{hubble:realm, hubble:lfnebulaestars, hubble:lfnebulaevelmag}, who claimed a Gaussian form. Various studies followed, which included data on dwarf galaxies in clusters \citep[e.g.][]{abell:clusmem} and the study of \citet{holmberg:groups} of galaxies in the field. Both of these argued for a power-law faint-end slope to the overall LF, as opposed to a Gaussian form.

In the mid 1970s it was suggested \citep{schechter:lf} that the optical LF can be approximated by the function \begin{equation} \label{eq: Schechter} \phi(L)~dL = \phi ^* \left( \frac{L}{L^*} \right) ^{\alpha} \mrm{exp} \left( - \frac{L}{L^*} \right) d \left( \frac{L}{L^*} \right) , \end{equation} where $\phi^*$ is the normalisation, $L^*$ is the characteristic luminosity above which the function has an exponential cutoff and $\alpha$ is the value of the power-law slope below $L^*$. The Schechter function, as it is now known, has been used for most subsequent characterisations of the universal LF. The function is motivated by self-similar gravitational condensation of structures \citep{press:structfmn}, except that $\alpha$ is not a fixed parameter.

Until recently the study of the LF, particularly for field galaxies as opposed to those in clusters, has been hampered by the lack of large samples. Few galaxies had redshifts and the photometry was from photographic plates, resulting in a large variation in the measured LF between different surveys. In the past decade data from large redshift surveys, such as the 2dF Galaxy Redshift Survey \citep[2dFGRS,][]{colless:2df,colless:2dffinal} and the Sloan Digital Sky Survey \citep[SDSS,][]{york:sdss} have become available, greatly improving the situation, although the LFs are still not in perfect agreement --- for example, \citet{liske:mgclf} gives revisions to the normalisations of the LF in various surveys based on the deeper wide-field imaging obtained by the Millennium Galaxy Catalogue (MGC), which they also describe, and recent comparisons of LFs such as \citet{driver:beyondlf} conclude that the variations between surveys are dominated by the systematic errors, particularly at the faint end. The differences are most likely due to surface brightness selection effects.

Whilst the universal LF is reasonably well constrained, it is well-known that the LF varies as a function of intrinsic galaxy properties. Thus the universal LF needs to be augmented by a description of this variation. As above, the lack of data until recently has hampered such studies, but with the large sample sizes (2dFGRS and SDSS) and five band CCD photometry (SDSS) now available, detailed divisions of the galaxies into samples which are still statistically significant is now possible. One can also apply fairly strict sample cuts to minimize biases and still retain large samples.

Previous studies of non-universal LFs have included 1) those focusing on specific environments, particularly clusters as the galaxies are all at approximately the same distance and in a narrow field of view, 2) those focussing on specific types of galaxies, the large number of studies which divide the general LF according to some criterion and fit functions to each bin, and 3) those which study the LF bivariate with another parameter, either binned or as an analytical function. Recent examples, some of which are bivariate, are \citet{popesso:clusterlf} for clusters, \citet{mercurio:superclusterlf} for superclusters, \citet{hoyle:voidlf} for voids, \citet{dejong:sbsdmlf} and \citet{delapparent:spirallfevoln} for spirals, and \citet{reda:earlylf}, \citet{stocke:earlytypelf} and \citet{dejong:bdearly} for ellipticals.

Various theories and simulations exist pertaining to the origin of the LF, tied in with the physics of galaxy formation. Examples are \citet{benson:lfshape} who begin with the mass function of dark matter haloes in the $\Lambda$CDM cosmology and add gas cooling, photoionization, feedback (e.g. from supernovae), galaxy merging and thermal conduction. \citet{mo:lfenvt} assume that the segregation of the galaxy population by environment is due to that of halo properties and use the halo occupation distribution to give predicted LFs by environment. \citet{cooray:lstar} reconstructs the Schechter form of the LF using empirical relations between the central galaxy luminosity and halo mass, and the total galaxy luminosity and halo mass.

Further details of the LF are in various reviews, for example \citet*{binggeli:lf} and subdivided by morphological type in \citet{delapparent:morphlf}.

In this paper the bivariate LF is studied in bins using the nonparametric stepwise maximum likelihood method of \citet*{efstathiou:swml} and two-dimensional analytical functions are fitted to the results. Previous studies that have done this include \citet{choloniewski:bivlf} subdivided by radius, \citet{sodre:esolvbivlf} subdivided by galaxy diameter and \citet{cross:ldens}, \citet{cross:bbd} and \citet{driver:mgcbbd} subdivided by absolute effective surface brightness. The latter quantity is also known as the bivariate brightness distribution.

\citet{ball:ann} showed that morphological types can be reliably assigned to galaxies in the SDSS using artificial neural networks (ANNs) provided a representative training set is available. Here types are assigned to the resolution E, S0, Sa, Sb, Sc, Sd and Im for 37,047 galaxies. In Ball et al. (2006, in preparation, hereafter B06) we use these types to study the variation of galaxy morphology with environment and compare the results to those for colour.

The bivariate LFs are computed for galaxies in the New York Value-Added Galaxy Catalogue \citep[VAGC,][]{blanton:vagc}, based on the Sloan Digital Sky Survey Data Release 4 \citep[DR4,][http://\-www.\-sdss.\-org/\-dr4]{adelmanmccarthy:dr4} for a range of galaxy properties, including the morphological type described, inverse concentration index, S\'ersic index, absolute effective surface brightness, eClass spectral type, reference frame colours, absolute 90\% radius, stellar mass and galaxy environment. The galaxy samples are flux limited to $r<17.77$, with the exception of the morphological type, which is limited to $r<15.9$.

Here the LF is only considered in the wavebands of the SDSS, but there are many studies in other wavebands, including the ultraviolet \citep[e.g.][]{baldry:ulf, budavari:galexlf, wyder:galexuvlf}, near-infrared K band \citep[e.g.][]{loveday:kbandlf,cole:nirlf,kochanek:kbandlf,eke:kband,jones:6dflf}, radio \citep[e.g.][]{sadler:radiolf} and x-ray \citep[e.g.][]{miyaji:xraylf, ranalli:xraylf}.

\citet{blanton:dr1lf} and \citet{loveday:lfevol} show that the LFs in the SDSS DR1 \citep{abazajian:dr1} to $z \simlt 0.3$ are consistent with evolution in the galaxy population over this redshift range. \citet{loveday:lfevol} shows that even to a redshift of 0.15 evolution in the number density occurs, but that it is only marginally significant in DR1. Here, because we are measuring and fitting many different bivariate functions, we do not attempt to account for evolution, but restrict the sample to $z \leq 0.15$ to minimize its effect.

Throughout, the standard spatial geometry is assumed, with Euclidean space, $\omegam = 0.3$ and $\omegal = 0.7$. For compatibility with previous studies, the dimensionless Hubble constant, $h = H_0/100 \kmsp$, is set to 1. 

\section{Data} \label{sec: data} 

The SDSS is a project to map $\pi$ steradians of the northern galactic cap in five bands ($u$, $g$, $r$, $i$ and $z$) from 3,500--8,900 \AA. This will provide photometry for of order $5 \times 10^{7}$ galaxies. A multifibre spectrograph will provide redshifts and spectra for approximately $10^{6}$ of these. A technical summary of the survey is given in \citet{york:sdss}. The telescope is described in \citet{gunn:sdsstelescope}. The imaging camera is described in \citet{gunn:sdsscamera}. The photometric system and calibration are described in \citet{fukugita:sdssphotometry}, \citet{hogg:sdssmt}, \citet{smith:ugrizstars}, \citet{ivezic:sdssdata} and \citet{tucker:calibration}. The astrometric calibration is in \citet{pier:sdssastrometry} and the data pipelines are in \citet{lupton:sdssphoto}, \citet{lupton:deblender} for the deblender, Frieman et al. and Schlegel et al. (in preparation).

The targeting pipeline chooses targets for spectroscopy from the imaging. A tiling algorithm \citep{blanton:tiling} then assigns the spectroscopic fibres to the targets, the main source of incompleteness being the minimum distance of 55 arcsec between the fibres. This causes about 6\% of galaxies to be missed, and these will be biased towards regions with a high surface density of galaxies. The algorithm gives a more uniform completeness on the sky than a uniform tiling by taking into account large scale structure, but some bias is still present.

The SDSS galaxies with spectra consist of a `Main', flux-limited sample, with a median redshift of 0.104 \citep{strauss:mainsample}, a luminous red galaxy sample (LRG), approximately volume-limited to $z \approx 0.4$ \citep{eisenstein:lrgsample} and a quasar sample \citep{richards:qsosample}. The limiting magnitude for the Main spectra is $r < 17.77$, which is substantially brighter than that for the imaging so the redshift completeness is almost 100\%. A typical signal-to-noise value is $>4$ per pixel and the spectral resolution is 1800. The redshifts have an RMS accuracy of $\pm 30 \kms$.

This paper uses galaxies from the DR4 version of the VAGC, which is described at http://\-wassup.\-physics.\-nyu.\-edu/\-vagc/. \citet{blanton:vagc} describes the DR2 \citep{abazajian:dr2} version. The catalogue is a publicly-available set of FITS files containing the Princeton reductions (http://\-photo.\-astro.\-princeton.\-edu; http://\-spectro.\-astro.\-princeton.\-edu) of the raw SDSS data with numerous additional derived quantities and matches to other surveys. The Princeton reductions are designed to improve on the original SDSS pipeline reductions in the publicly available DR4 Catalogue Archive Server (CAS; http://\-cas.\-sdss.\-org/\-astro/\-en) and are used throughout unless otherwise stated. The set of objects contained is designed to match the SDSS DR4 and contains galaxies which either match a slightly more inclusive version of the Main sample criteria, are within 2 arcsec of a Main, LRG or quasar target from the version of the photometry used for the targeting, or are within 2 arcsec of a hole drilled in an SDSS spectroscopic plate. The catalogue provides 6,851 square degrees of imaging coverage and 4,681 square degrees of spectroscopic coverage. The raw catalogue contains 1,223,536 objects, 722,866 of which have spectra.

The VAGC units differ from the CAS data in some quantities: the object brightnesses are given as fluxes in nanomaggies (where $1~{\mrm{maggie}} = 0 \magarcsec$); the errors are inverse variance and the Petrosian radii are in pixels. These are converted to the units used in which the Main Galaxy Sample is defined, i.e. magnitudes, $1\sigma$ errors and arcsec. The pixels are converted to arcsec using the pixscale values given in the full VAGC calibObj photometry outputs. These are all close to 0.396 arcsec pixel$^{-1}$.

From the catalogue, the main file containing the imaging data, object\_sdss\_imaging.fits, was chosen as a base, and the rest of the required parameters for each object were obtained by matching the other datafiles to this one using a method appropriate for each parameter and file in question. The matches are described further below.

The magnitudes are corrected for galactic reddening using the usual corrections derived from the maps of \citet*{schlegel:dustmaps}. These are of order $-0.1 \mgn$, with a mean value of $-0.14 \mgn$ and $1\sigma$ variation of $0.11 \mgn$. The minimum and maximum values are $-0.02 \mgn$ and $-1.04 \mgn$ respectively, both in the $u$ band.

K-corrections are applied using version 4.1.14 of the code described in \citet{blanton:kc}. This provides a separate K-correction for each individual galaxy, but does not take into account galaxy evolution or dust. The K-corrections were made to a band-shift corresponding to a redshift of 0.1. Most of the datasets have approximately this mean. One could attempt to K-correct to the band-shift for the mean of each dataset, however, this would affect the dataset chosen and so this is not attempted here.

\subsection{Galaxy Samples and Properties} \label{sec: properties}

The SDSS DR4 imaging outputs were used directly as supplied in the VAGC file object\_sdss\_imaging.fits and cross-matched from the full outputs available in the calibObj files. The DR4 spectroscopic outputs were used similarly from object\_sdss\_spectro.fits. 

The absolute magnitude range used for each bivariate LF was $-24 < \mrlogh < -15$ in 18 bins of $0.5 \mgn$, with two exceptions: the JPG morphological type (see below) was binned in 12 bins of $0.5 \mgn$ for $-24 < \mrlogh < -18$, and the environmental density was in 22 bins for $-24.03 < \mrlogh < -19.63$, due to its restricted redshift range (see below). Details of each bivariate LF are given in Table \ref{table: bivlfs}.

The upper limit of approximately 40 evenly spaced bins in the second parameter was chosen to give a reasonable execution time. The extent of the bins is based on visual inspection of the distributions of the input parameters, but the value of the LF in each bin is independent of the others.

Our analysis allows for the sampling rate of the dataset, which is the ratio of the number of galaxies which have spectra taken to those meeting the criteria for spectra to be taken. The completeness maps as a function of imaging-only parameter pairs are given in \S \ref{sec: LF Results}. The overall completeness is estimated to be 93\%. 6\% of galaxies are missed due to the minimum 55 arcsec separation between adjacent spectroscopic fibres on a plate, and the Main Galaxy Sample is about 99\% complete overall \citep{strauss:mainsample}. However, this completeness is in terms of those objects that are targeted for spectroscopy in the first place, and in particular is subject to an explicit Petrosian 50\% light surface brightness cut of $\sbrapp < 24.5 \magarcsec$ for all objects. \citet{blanton:lflowl} show that the faint end is incomplete. We therefore calculate the sampling rate for each bin in the bivariate LF and correct for this in the estimation of the LF (see \S \ref{sec: BivLFs}).

Because we calculate the sampling rate, the galaxies used in the process of obtaining our results do not necessarily have spectra. We therefore apply the set of recommended photometric flags described at http://\-www.\-sdss.\-org\-/dr4 to generate a clean sample. We also require the resolve\_status flag to be survey\_primary and the vagc\_select flag to be main.

For spectra, as mentioned, the VAGC Main-like sample cuts are slightly more inclusive than the SDSS Main sample, in particular allowing in some binary stars by the removal of the cut for small bright objects. Here we reapply the Main criteria as given in \citet{strauss:mainsample}. For objects with spectra we also require the specprimary flag to be set, the primtarget flag to be set to either galaxy, galaxy\_big, or galaxy\_bright\_core, the progname flag to be main, the zwarning flag to be 0 and the platequality flag to be good. The VAGC does not contain a redshift confidence parameter equivalent to the zConf in the CAS, but bad spectra and redshifts are excluded by zwarning.

In the optical, many spiral galaxies that are edge-on or near to this suffer internal extinction and reddening due to dust. Here we attempt to minimize this effect by restricting the sample to galaxies with an axis ratio of less than that for an E7 elliptical galaxy. Galaxies with larger axis ratios than this are likely to be edge-on or close to edge-on spirals. Because the radial light profiles available from the SDSS and VAGC are either axisymmetric or specific to certain types of galaxies (de Vaucouleurs or exponential), we use the $25 \magarcsec$ isophotal major and minor axes isoA and isoB in $r$ and restrict the sample so that ${\mrm{isoA}} / {\mrm{isoB}} < 10/3$. This excludes 30,407 (2.5\%) of the galaxies. Fig. 4 of \citet{vincent:shape} shows that the vast majority of the bright galaxies removed are those with an exponential or mixed exponential and de Vaucouleurs profile. Very few galaxies with a pure de Vaucouleurs profile are removed. At faint magnitudes similar numbers of galaxies are removed for both profile types, so many more faint ellipticals are removed than bright. Here `bright' depends on the profile type, varying from $\mrlogh < -21.1$ for de Vaucouleurs to $\mrlogh < -19.7$ for exponential.

A potential alternative measure is the axis ratio derived from the adaptive moments \citep{stoughton:edr}. However, \citet{kuehn:shape} show that this would be affected by seeing for most of the galaxies in our sample due to their small sizes and consequent poor resolution. As our purpose is simply to exclude very elongated objects we use the isophotal cut. A similar example is \citet{zibetti:edgeon}, who require ${\mrm{isoB}} / {\mrm{isoA}} < 0.25$ in $g$, $r$ and $i$, and $a > 10~{\mrm{arcsec}}$ for their sample of edge-on galaxies.

The VAGC also contains eyeball quality checks of spectra for 17,422 objects, mostly bright ($\mrlogh < -15$) or of low redshift ($z < 0.01$). The sample is not complete in any particular sense, but it does exclude numerous spurious objects such as bad deblends of large galaxies, HII regions, clearly incorrect classifications and so on. We require that the quality flag be set to done with no other bit set, unless use\_anyway is set. The requirement noticeably reduces the numbers of objects in some of the sparsely populated outlying bins of the LF.

Besides the VAGC, we use data from five other sources: 1) the Japan Participation Group (JPG) catalogue of eyeball-classified galaxy morphologies, based on the SDSS Early Data Release \citep[EDR,][]{stoughton:edr}, 2) the neural network morphological types of \citet{ball:ann}, 3) the public SDSS DR4 CAS, 4) the Max Planck/Johns Hopkins (MPA/JHU) catalogues for galaxies and AGN at http://\-www.\-mpa\--garching.\-mpg.\-de/\-SDSS and 5) the Pittsburgh-Carnegie Mellon Value-Added Catalogue (VAC; http://\-nvogre.\-phyast.\-pitt.\-edu/\-dr\_value\_added) for galaxies. The CAS gives the eClass spectral type, the MPA/JHU the stellar mass and the VAC the environmental density.

\begin{table*} \centering
\caption{Parameters and ranges used in the bivariate luminosity distributions. See text for further details on the parameters and data sources. All data are based on the SDSS DR4, with the exception of the JPG catalogue, which is from the EDR. ${\mrm{N_{gal}}}$ is the number of galaxies, to the flux limit, with spectra and selected for the luminosity function respectively. The overall sampling rate is ${\mrm{N_{gal_{spectro}}}} / {\mrm{N_{gal_{imaging}}}}$. $\alpha$, $M^*$ and $\phi*$ are the best-fitting faint-end slope, characteristic magnitude and normalisation respectively for the overall best-fit Schechter function for each LF. \label{table: bivlfs}}
\vspace{0.25cm}
{\tiny
\begin{tabular}{lccccccccclll} \hline
Parameter &Symbol &Min &Max &${\mrm{N_{gal_{imaging}}}}$ &${\mrm{N_{gal_{spectro}}}}$ &${\mrm{N_{gal_{LF}}}}$ &$\alpha$ &$M^*$ &$\phi*$ &Photometry &Type &Data source\\
\hline
overall                         &LF          &-24   &-15   &489,123 &238,737 &237,337 &-1.37 &-20.49 &0.0159 &Princeton &S\'ersic  &NYU VAGC\\
morphological T type            &$\tjpg$     &-0.5  & 6.5  &1,150   &1,019   &998     &-1.30 &-20.58 &0.0098 &CAS       &Petrosian &JPG Catalogue\\
ANN T type                      &$\tann$     &-0.5  & 6.5  &35,604  &21,352  &20,891  &-0.95 &-20.34 &0.0117 &CAS       &Petrosian &ANN morphology\\
inverse concentration index     &$\ciinv$    & 0.1  & 0.7  &489,102 &238,737 &237,337 &-1.14 &-20.46 &0.0141 &Princeton &S\'ersic  &NYU VAGC\\
log(S\'ersic index)             &$\logn$     &-0.6  & 0.75 &448,788 &223,700 &222,458 &-1.13 &-20.44 &0.0135 &Princeton &S\'ersic  &NYU VAGC\\
abs. eff. SB / $\magarcsec$     &$\sbr$      &17    &24.5  &487,789 &238,256 &236,857 &-1.24 &-20.49 &0.0136 &Princeton &S\'ersic  &NYU VAGC\\
reference frame $u-r$           &$u-r$       & 0    & 5.5  &413,792 &237,732 &237,157 &-1.23 &-20.48 &0.0118 &Princeton &S\'ersic  &NYU VAGC\\
reference frame $g-r$           &$g-r$       &-0.25 & 1.5  &414,398 &237,881 &237,265 &-1.22 &-20.48 &0.0119 &Princeton &S\'ersic  &NYU VAGC\\
reference frame $r-i$           &$r-i$       &-0.5  & 1    &414,377 &237,882 &237,301 &-1.23 &-20.48 &0.0118 &Princeton &S\'ersic  &NYU VAGC\\
reference frame $r-z$           &$r-z$       &-1    & 1    &414,273 &237,875 &237,175 &-1.22 &-20.48 &0.0119 &Princeton &S\'ersic  &NYU VAGC\\
log(90\% radius) / $\kpc$       &$\absrad$   &-0.3  & 1.6  &487,780 &238,256 &236,851 &-1.33 &-20.55 &0.0127 &Princeton &S\'ersic  &NYU VAGC\\
eClass                          &$\eclass$   &-0.6  & 1    &273,582 &219,171 &218,000 &-1.24 &-20.73 &0.0061 &CAS       &Petrosian &SDSS CAS\\
log(stellar mass) / $\msun$     &$\mstellar$ & 6    &13    &288,354 &231,613 &230,246 &-1.37 &-20.80 &0.0056 &CAS       &Petrosian &MPA/JHU\\
log(surface density) / $\dens$  &$\Sigma_5$  & 0    & 1.2  &213,343 &25,440  &8,003   &-2.11 &-22.56 &0.0002 &CAS       &Petrosian &Pitt/CMU VAC\\
\hline
\end{tabular}
}
\end{table*}

\subsection{Petrosian and S\'ersic Magnitudes} \label{sec: magnitudes}

Petrosian magnitudes measure a constant fraction of the total light, in a model-independent manner. They are available in the SDSS in a modified form from that introduced by \citet{petrosian:petromag}. The Petrosian flux is given by \begin{equation} F_{\rmn{P}} \equiv \int_{0}^{N_{\rmn{P}}r_{\rmn{P}}} 2\pi r'dr'I(r') \end{equation} where $r_{\rmn{P}}$ is the Petrosian radius, which is the value at which the Petrosian ratio of surface brightnesses \begin{equation} R_{\rmn{P}}(r) \equiv \frac{\int_{0.8r}^{1.25r} 2\pi r'dr'I(r')/[\pi (1.25^2-0.8^2)]}{\int_{0}^{r}2\pi r'dr'I(r')/(\pi r^2)} \end{equation} has a certain value, chosen in the SDSS to be 0.2. The number $N_{\rmn{P}}$ of Petrosian radii within which the flux is measured is equal to 2 in the SDSS. Further details are given in \citet*{lupton:asinhmag} and \citet{stoughton:edr}.

The Petrosian radii are not seeing corrected, which causes the surface brightness and concentration to be underestimated for objects of size comparable to the PSF. However the seeing effect is not yet quantified, and there are other approximations such as using a circular as opposed to elliptical aperture and not correcting for dust obscuration. As described in \citet{stoughton:edr} the Petrosian aperture is not missing much flux compared to an ideal galaxy light profile, the amount missing being about 20\% for a de Vaucouleurs profile and only 1\% for an exponential profile. The effect of seeing, which would make the profiles tend towards a PSF, for which 5\% of the light is lost, is also small for galaxies in the main sample. \citet{blanton:dr1lf} show that the resulting luminosity density in $r$ is very similar to that from the S\'ersic profile, the difference being $j_{0.1}(\mrm{Sersic}) = j_{0.1}(\mrm{Petrosian}) - 0.03$. They suggest that the similarity shows that the true luminosity density is also of a similar value.

The VAGC also contains S\'ersic fits to the galaxy light profiles. The S\'ersic profile \citep{sersic:australes,graham:sersic} is obtained by generalising the de Vaucouleurs profile \citep{devaucouleurs:devprofile} to have index $\frac{1}{n}$ instead of $\frac{1}{4}$, giving \begin{equation} I(r)=I_0~{\mathrm{exp}}\{-b_n[(r/r_{\mathrm{e}})^{1/n}]\}, \label{eq: Sersic} \end{equation} where $b_n$ is such that half the total luminosity is within $r_e$. The fitting procedure used in the VAGC is described in \cite{blanton:vagc} and in more detail in the Appendix of \cite{blanton:envtbbprops}. The axisymmetric S\'ersic profile is fitted to the azimuthally averaged galaxy light profile available from the SDSS database. The profiles are corrected for seeing, which is modelled using three axisymmetric Gaussians. The de Vaucouleurs and exponential profiles correspond to $n=4$ and $n=1$ respectively.

\citet{graham:totalmag} show the discrepancy between the Petrosian and S\'ersic magnitudes. In particular, the difference depends on the shape of the galaxy light profile.  Their table 1 for the SDSS Petrosian aperture shows that for inverse concentration index ($\ciinv$; \S \ref{sec: dr4}), the difference is negligible above $\ciinv = 0.35$, but below this increases to, where $\mu_e$ is the total effective half light surface brightness, $\mu_e - \sbrapp = 0.52$, 1.09 and 1.74 $\magarcsec$ for $\ciinv = 0.30$, 0.28 and $0.27$ respectively for mean half-light surface brightness, and $m_{\mrm{Petrosian}} - m_{\mrm{S\acute{e}rsic}} = 0.22 \mgn$, $0.38 \mgn$ and $0.54 \mgn$ in magnitude over the same range. Thus only a few percent of our galaxies are affected significantly. Their fig. 6 shows that the VAGC S\'ersic fits are consistent with the expected differences. The $\ciinv$ of 0.28 corresponds to a S\'ersic index of 6.

Although the S\'ersic index is an improvement in that it fits a more accurate radial light profile to a galaxy and is seeing-corrected, there are still some biases present. In their fig. 9, \citet{blanton:vagc} show the residuals of their S\'ersic fits for 1,200 simulated galaxies. As a function of $n$, the value of $n_{out} - n_{in}$, where $n_{out}$ is the fitted value of $n$, decreases monotonically from less than -0.05 at $n=1$ to -0.1, -0.25, -0.5 and -0.65 at $n=2,3,4$ and 4.5 respectively. Thus a de Vaucouleurs galaxy ($n = 4$) is assigned an index of $n \sim 3.5$ and the indices are all systematically underestimated. The error of 0.5 is comparable to the bin size in the same area of our bivariate luminosity-S\'ersic index distribution. The range of the error as shown by the quartiles also broadens from around 0.1 to 0.5 over the same range. The fitted 50\% light radius and flux decrease similarly over their range to around 0.8 and 0.9 of their true values. Also, the fitted radius and flux decrease with $n$, the fitted flux with radius and the fitted radius with flux. However, the fitted $n$ does not decrease with radius and flux. The overall performance for $n$ is characterized as good because the bias is similar to the uncertainty and comparable but opposite to that from the assumption of axial symmetry. The overall fits are thought to be approximately correct for S\'ersic-shaped galaxies and supply a seeing-corrected estimate of size and concentration for the others.

Here we investigated the LFs for both Petrosian and S\'ersic profiles. The LFs are generally similar and because the samples used are based on the SDSS Main Galaxy Sample, which is based on Petrosian magnitudes, we present the results based on those magnitudes. The derivation of a fully S\'ersic-based sample is beyond the scope of this paper.

\subsection{Neural Network Morphological Types} \label{sec: nn types}

1,875 SDSS galaxies to $r<15.9$ in the SDSS EDR have been classified into morphological types by \citet{nakamura:morphlf}, forming the JPG catalogue. The system used was a modified version of the T-type system \citep{devaucouleurs:ttype}, with the types being assigned in steps of 0.5. The corresponding Hubble types are E=0, S0=1, Sa=2, Sb=3, Sc=4, Sd=5 and Im=6.

We have previously shown \citep{ball:ann} that artificial neural networks \citep[e.g.][]{bishop:ann,lahav:annmethods} are able to assign types to galaxies in the SDSS with an RMS accuracy of 0.5 on this scale, using the JPG catalogue as a training set. This is the same as the spread between the human-classified types by the members of the JPG team.

We used the same procedures here, updated for DR4 over DR1 and with some modifications. The DR4 CAS was matched to the JPG catalogue using a tolerance of 2 arcsec and a flux limit of $r<15.9$. Unassigned types ($-1$) and galaxies flagged as being likely to have bad photometry were removed as before. The clean photometry flags described above were also applied here. After these cuts 1,290 galaxies remained, 1,131 with spectroscopic redshifts. The training and test sets were 645 each with no overlap, from which, as before, galaxies with severely outlying parameters and targets ($>10\sigma$ from the mean value for the parameter, generally indicative of a measurement error) were iteratively removed for each parameter in turn. The network trained on the 645 was applied to the DR4 sample using the CAS photometry. One could also retrain on all 1,290 for the final network to be applied to the sample, but it would make very little difference.

It was also found that when trained on the full set of 29 parameters in \citet{ball:ann} the types were biased towards early type with increasing redshift to a greater extent than in the JPG catalogue. However, when one trains on the purely morphological subset of the parameters (all those except the magnitudes and colours), the bias is similar to the JPG, reflecting the flux-limited nature of the sample, but the RMS variation between network type and target type is not significantly larger. Hence the latter set is used as the training set here.

The network also has a tendency to `avoid the ends of the scale', resulting in few very late type galaxies and a slight bias away from early type galaxies. This is shown in Fig. 1 of \citet{ball:ann}. This is likely due to the severe lack of galaxies in our training set of type $T=5$ or later compared to the earlier types, and the importance of the concentration index in the morphological training set, which shows a similar deviation away from the ends of the scale for both early and late types.

We restrict the absolute magnitude range of the morphologies presented to $\mrlogh < -18$. This is due to a lack of galaxies in the training data at fainter magnitudes than this. Thus the neural network types are not extrapolated from the training set in apparent or absolute magnitude.

\subsection{SDSS DR4 VAGC Data} \label{sec: dr4}

The bivariate LF parameters described in this section are the inverse concentration index, S\'ersic index, absolute effective surface brightness, reference-frame colours and absolute radius. The morphological parameters are measured in the $r$ band, since this band is used to define the aperture through which Petrosian flux is measured for all five bands.

The inverse concentration index $\ciinv$ is $R_{50}/R_{90}$ where $R_{50}$ and $R_{90}$ are the radii within which 50 and 90 per cent of the Petrosian flux is received. The inverse is used because it has the range 0--1.

The S\'ersic index $n$ is monotonically related to the concentration \citep[e.g.][]{graham:totalmag}, if the latter is measured using S\'ersic radii. Here we use Petrosian circular radii and for the bivariate luminosity-S\'ersic index distribution we use $\logn$, as the index is in the exponent in the equation defining the S\'ersic profile (equation \ref{eq: Sersic}).

The absolute effective surface brightness used here is given by \begin{equation} \sbr = m_{r} + 2.5\mrm{log}~(2 \pi R_{\mathrm{50}}^2) - 10 {\rmn{log}} (1+z) - K, \end{equation} where $m_{r}$ is the $r$ band magnitude and $K$ is the K-correction. The evolutionary correction term is set to zero.

The reference frame colours are taken from the K-corrected values described above. All ten colours from $u$, $g$, $r$, $i$ and $z$ were investigated.

The absolute radius of a galaxy at redshift $z_{\mrm{gal}}$ is calculated from the apparent Petrosian radius using the $\omegam = 0.3$, $\omegal = 0.7$, $h=1$ cosmology. It is also shown logarithmically and is given by \begin{equation} \absrad = \frac{R_{90}~d(z)}{1+z}, \end{equation} where $d$ is the relativistic coordinate distance \begin{equation} d = \frac{c}{H_0} \int_{0}^{z_{\mrm{gal}}} \left[ (1-\omegal)(1+z)^3 + \omegal \right]^{-0.5}. \end{equation}

\subsection{Outputs from additional SDSS Catalogues} \label{sec: additional}

The additional catalogues used are all based on SDSS DR4 and use the same Euclidean space, $\omegam = 0.3$, $\omegal = 0.7$, as used here. The quantities all require spectra to be calculated and are therefore matched using the unique spectroscopic identification, both in object and time of observation, provided by the quantities plate, mjd and fiberid.

The eClass \citep{connolly:orthogonal,connolly:eclass,yip:eclass} is a continuous one-parameter type assigned from the projection of the first three principal components (PCs) of the ensemble of SDSS galaxy spectra. The locus of points forms an approximately one dimensional curve in the volume of PC1, PC2 and PC3. This is a generalization of the mixing angle $\phi$ in PC1 and PC2 \begin{equation} \phi = {\rm tan}^{-1} \left( \frac{a_2}{a_1} \right) , \end{equation} where $a_1$ and $a_2$ are the eigencoefficients of PC1 and PC2. The range is from approximately $-0.6$, corresponding to early type galaxies, to 1, late type. The eClass is also robust to missing data in the spectra used for its derivation, and is almost independent of redshift. The quantity is not given in the VAGC and so was matched to the public SDSS DR4 data from the CAS.

The MPA/JHU value-added catalogues at http://\-www.\-mpa-garching.\-mpg.\-de/\-SDSS include the quantity rml50, the logarithm of the median dust-corrected stellar mass in solar units. It is calculated from the stellar mass to light ratio predicted by a large library of models of star formation history and is described further in \citet{kauffmann:masssfh}.

The VAC contains a measure of galaxy environment as the parameter density\_\-z005095. This is the distance to the $N$th nearest neighbour within $\pm 1000 \kms$ in redshift for $0.053 < z < 0.093$. The $\pm 1000 \kms$ is used to minimize contamination from interlopers. Galaxies for which the survey edge is reached before the $N$th nearest neighbour are excluded to avoid downward bias in the estimated densities near the survey edges. Here the value of $N$ used is 5, following \citet{balogh:bimodallfenvt} who choose this value to approximate \citet{dressler:morphdensity} who uses $N=10$ before correction for superimposed galaxies. The surface density of galaxies is then given by $\Sigma_N = N/\pi d_N^2$. The nearest neighbour must be of magnitude $r < 17.7$ at $z=0.093$, which corresponds to $\mrlogh < -19.63$. This cut ensures a uniform density measurement over the redshift range of the sample rather than the density decreasing with redshift due to missing faint galaxies. The value of $h$ used in preparing this catalogue was $h=0.7$ so the densities are given here in units of $\dens$.

\section{Bivariate Luminosity Functions} \label{sec: BivLFs}

Unlike the monovariate LF in which the Schechter function is often used, there is no common functional form for the bivariate LF, so a nonparametric estimator is chosen. We use the popular stepwise maximum likelihood (SWML) method of \citet{efstathiou:swml}, following the extension to a bivariate distribution by \citet{sodre:esolvbivlf}. The maximum likelihood method has well defined error properties \citep{kendall:stat} and in the comparison between LF estimators of \citet{takeuchi:lfests} the SWML was shown to be a good estimator.

The SWML gains independence of inhomogeneities in the galaxy distribution by assuming that this is the case via the universal form: $n(L,\mrm{\bf x}) = \phi(L)\rho(\mrm{\bf x})$. This means that the $\phi(L)$ shape is independent of its normalisation, which then has to be found separately. The data must also be binned over the ranges of the parameters considered.

Here the bivariate LF is given by $\psi(L,X)$ or $\psi(M,X)$ where $X$ is the second parameter.

For a galaxy which is observable if its luminosity $L$ lies in the range $L_{\mrm{min}}$--$L_{\mrm{max}}$ and $X$ lies in the range $X_{\mrm{min}}$--$X_{\mrm{max}}$ (these limits in general being redshift dependent), the probability of seeing that galaxy with luminosity $L_i$ and $X=X_i$ at redshift $z_i$ is \begin{multline} p_i \propto \psi(L_i,X_i)~f(L_i,X_i) \\ \bigg / \int_{L_{\mrm{min}}(z_i)}^{L_{\mrm{max}}(z_i)} \int_{X_{\mrm{min}}(z_i)}^{X_{\mrm{max}}(z_i)} \psi(L,X)~f(L,X)~dL~dX , \label{eq: Probability} \end{multline} where $f$ is the completeness function, calculated for each bin in the bivariate LF. This automatically takes into account both the 6\% sampling incompleteness resulting from galaxies more closely spaced than the spectroscopic fibres, and the further 1\% from objects missed in the Main Galaxy Sample (see \S \ref{sec: properties}). It does not take into account any incompleteness in the imaging itself, which is discussed further in Appendix \ref{app: completeness maps} below.  The $X$ limits are a function of $z$ if the sample is explicitly or implicitly selected on $X$. The samples are not explicitly selected on $X$ if the quantity requires a spectrum.

The likelihood ${\mathcal L} = \prod p_i$ is then maximised with respect to $\psi(L,X)$. In practice, one maximises the log-likelihood $\mrm{ln}~{\mathcal L} = \sum \mrm{ln}~p_i$.

For the SWML method, the discrete version of this is used. $\psi(M,X)$ is parametrized as the number density of galaxies \begin{equation} \psi(M,X) = \psi_{jk} \qquad (j = 1 \ldots N_M;~k = 1 \ldots N_X) \end{equation} in $N_M$ and $N_X$ evenly spaced bins in absolute magnitude $M^-_j < M_j < M^+_j$ and $X^-_k < X_k < X^+_k$, where \begin{equation} M^{\pm}_j = M_j \pm \frac{\Delta M}{2} \end {equation} and \begin{equation} X^{\pm}_k = X_k \pm \frac{\Delta X}{2} . \end{equation} Magnitude bins are preferred to luminosity bins as the distribution is more evenly spread.

The likelihood is then given using the discrete equivalent of equation \ref{eq: Probability} \begin{multline} \mrm{ln}~{\mathcal L} = \sum_{i=1}^{N_g} \sum_{j=1}^{N_M} \sum_{k=1}^{N_X} W_{ijk}~\mrm{ln}~[\psi_{jk}~f(M_i, X_i)] \\ - \sum_{i=1}^{N_g} \mrm{ln} \left( \sum_{j=1}^{N_M} \sum_{k=1}^{N_X} H_{ijk}~\psi_{jk} \right) + \mrm{constant} , \label{eqn:like} \end{multline} where \begin{equation} W_{ijk} = \left\{ \begin{array}{ll} 1 & \mrm{if} - \Delta M/2 \leq M_i-M_j < \Delta M/2\\&\mrm{and} -\Delta X/2 \leq X_i-X_k < \Delta X/2,\\0 & \mrm{otherwise} \end{array} \right. \end{equation} and \begin{equation} H_{ijk} = \frac{1}{\Delta M \Delta X} \\ \times \int_{M^-}^{M'} \int_{X^-}^{X'} f(M,X)~dM~dX , \end{equation} where $M' = \mrm{max}[M^-, \mrm{min}(M^+,M^i_{\mrm{lim}})]$ and $X' = \mrm{max}[X^-, \mrm{min}(X^+,X^i_{\mrm{lim}})]$.

The constant in (\ref{eqn:like}) is fixed by using a Lagrangian multiplier $\lambda$: the constraint \begin{equation} g = \sum_j \sum_k \psi_{jk}  \Delta M \Delta X - 1 = 0 \end{equation} is applied and the new likelihood $\mrm{ln}{\mathcal L}^{'} = \mrm{ln}{\mathcal L} + \lambda g(\psi_{jk})$ is maximised with respect to $\psi_{jk}$ and $\lambda$, requiring $\lambda = 0$. 

The maximum likelihood estimate $\partial \mrm{ln} {\mathcal L}^{'} / \partial \psi_{jk} = 0$ is then \begin{equation} \psi_{jk} = \frac{\sum_{i=1}^{N_g} W_{ijk}}{\sum_{i=1}^{N_g}[H_{ijk} / \sum_{l=1}^{N_M} \sum_{m=1}^{N_X} \psi_{lm} H_{ilm} ]} , \end{equation} where $\psi_{lm}$ is from the previous iteration.

Following \citeauthor{efstathiou:swml}, the errors on the parameters are estimated using the fact that the MLM estimates $\psi_{jk}$ are asymptotically normally distributed with the covariance matrix \begin{equation} \mrm{cov} (\psi_{jk}) = \mathsf{I}^{-1}(\psi_{jk}), \end{equation} where $\mathsf{I}$ is the information matrix (their equation 2.13b).

For the normalisation, the estimator of the space density $\overline n$ of galaxies is \begin{equation} \label{eq: Normalisation} f\overline n = \sum_{i=1}^{N_{gal}}w(z_i) \bigg / \int_{z_{min}}^{z_{max}} S(z)w(z)~dV . \end{equation} The selection function $S$ from $L_1$ to $L_2$ is \begin{equation} S(z) = \sum_{L'_1}^{L'_2} \sum_{X'_1}^{X'_2} \psi(L,X) \bigg / \sum_{L_1}^{L_2} \sum_{X_1}^{X_2} \psi(L,X), \end{equation} where $L'_1 = \mrm{max}(L_{\mrm{min}}(z),L_1)$, $L'_2 = \mrm{min}(L_{\mrm{max}}(z),L_2)$ and similarly for $X$. The limits on the sums in the numerator depend on the K-correction, which is determined using the average SED for the sample. These limits will in general include partial bins with appropriate weighting due to the summation limits not falling at the edge of a bin.

We use the weighting function \begin{equation} w(z) = \frac{1}{[1+4\pi f\overline nJ_3(r_c)S(z)]} , \label{eq: Hamilton} \end{equation} where \begin{equation} J_3(r_c) = \int_{0}^{r_c} r^2\xi (r) dr , \end{equation} $r$ is the distance and $\xi (r)$ the real space galaxy two point correlation function. 

The variance in the correlation function can be estimated using a complex formula which minimizes the variance in the function and involves the three and four point correlation functions. These are more difficult to measure that the two point and require a large sample such as the SDSS. The weight function in equation \ref{eq: Hamilton} was found to be a good approximation by \citet{hamilton:cf} and is widely used. The function minimizes the variance in the estimate of the number density so long as $r_c$ is much less than the survey depth. Here $4\pi J_3 = 32000~h^{-3}~\mrm{Mpc^3}$, using the $\xi (r)$ from \citet{zehavi:edrclustering} which is given as $\xi (r) = (r/6.1 \pm 0.2~h^{-1}~{\rm Mpc})^{-1.75 \pm 0.03}$ for $0.1~h^{-1}~{\rm Mpc} \leq r \leq 16~h^{-1}~{\rm Mpc}$. The effect of using an incorrect value would be to change the normalisation but not the shape of the LF. For example, \citet{loveday:lfevol} uses the same values of $J_3$ and $\xi$ as here and finds that halving $J_3$ reduces the estimated density by 7\%. An update to \citet{zehavi:edrclustering}, \citet{zehavi:lcgcf}, has a similar $\xi$ value to that used here. 

Equation \ref{eq: Normalisation} is solved iteratively as it contains $w$, which contains $\overline n$. The minimum-variance $w(z)$ is described further in \citet{davis:cfadens} and \citet{hamilton:cf}.

\section{Results} \label{sec: LF Results}

\subsection{Morphological Type} \label{sec: LF T}

\begin{figure}
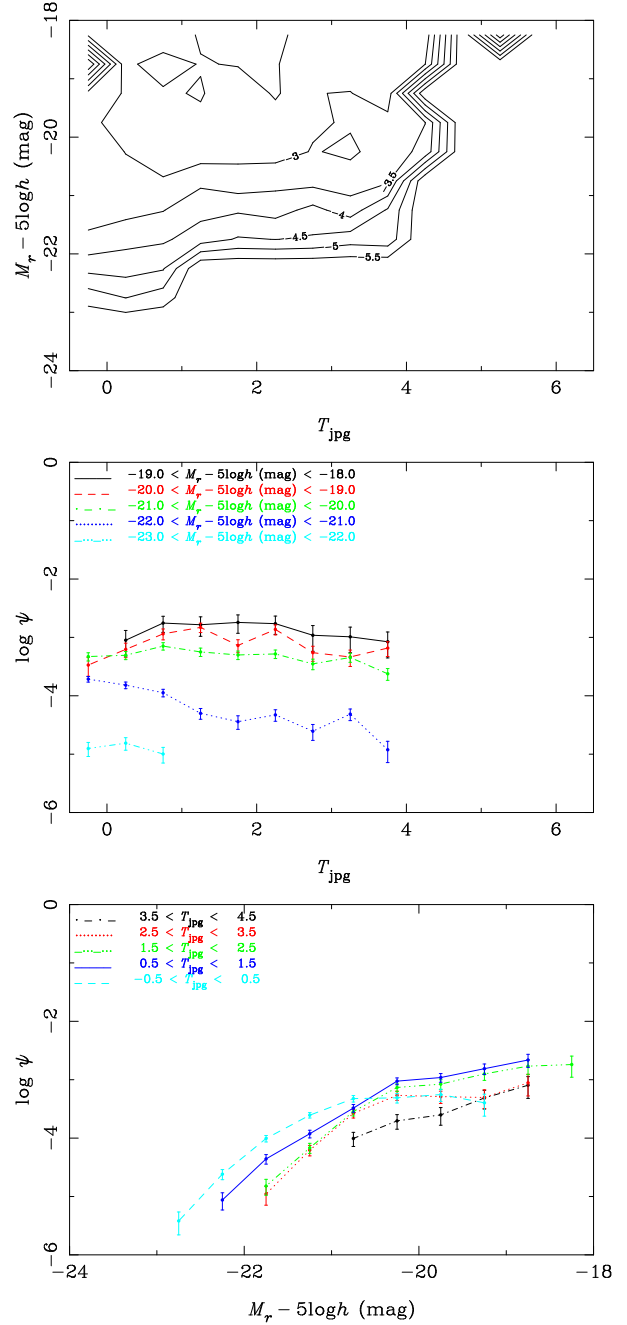
 \centering
\includegraphics[height=0.45\textwidth,angle=270]{figures/figure_01_panel1.eps}
\includegraphics[height=0.45\textwidth,angle=270]{figures/figure_01_panel2.eps}
\includegraphics[height=0.45\textwidth,angle=270]{figures/figure_01_panel3.eps}
\caption{LF bivariate with JPG catalogue morphological type $\tjpg$ for 998 galaxies. The contours show the value of $\mrm{log}_{10} \psi(M,X)$, the bivariate luminosity-$X$ distribution in each bin. The second panel shows vertical slices of $\psi(M,X)$ subdivided by $X$ for the ranges given and the lower panel shows the horizontal slices $\psi(M,X)$ subdivided by $M_r$. Although the contours show all galaxies, a point appears for the slice only if created by five or more galaxies, as below this threshold Poisson fluctuations dominate. In the rest of the LF plots (Figs. \ref{fig: ANN bivlf}--\ref{fig: Density bivlf}) this is increased to twenty (see text). Here $X$ is $\tjpg$. The errors are from the inverse of the information matrix (see text). Note that for clarity the axes in this figure and that for $\tann$ differ from subsequent figures due to the brighter cutoff in absolute magnitude. \label{fig: JPG bivlf}}
\end{figure}

\begin{figure}
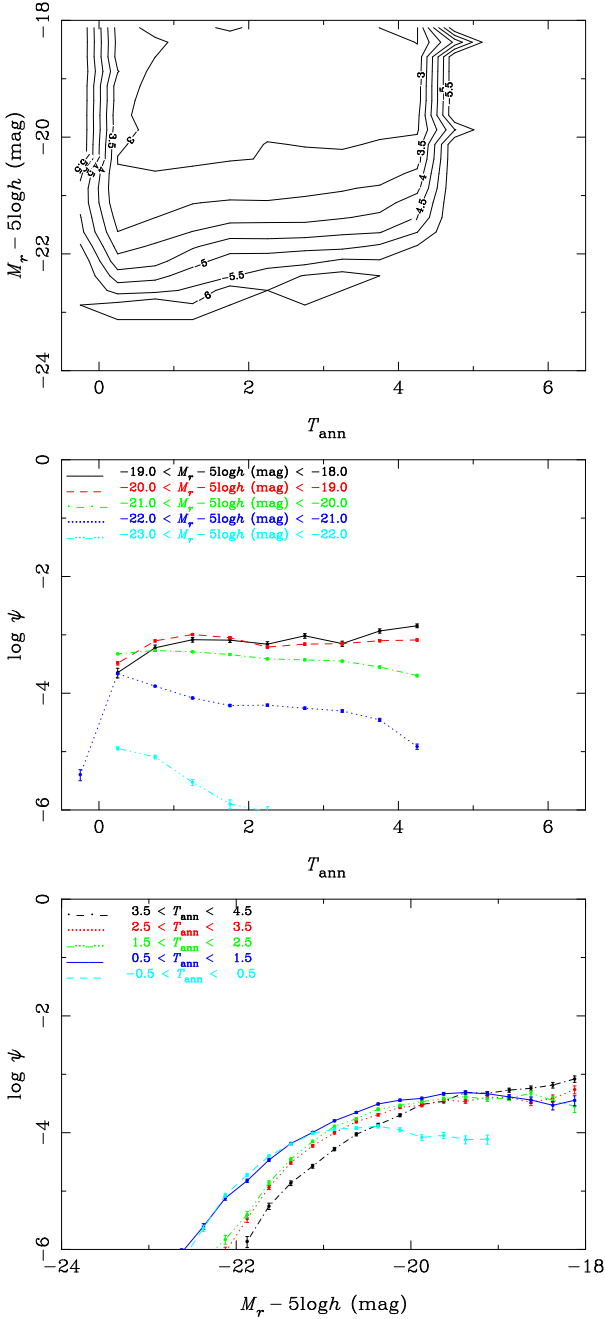
 \centering
\includegraphics[height=0.45\textwidth,angle=270]{figures/figure_02_panel1.eps}
\includegraphics[height=0.45\textwidth,angle=270]{figures/figure_02_panel2.eps}
\includegraphics[height=0.45\textwidth,angle=270]{figures/figure_02_panel3.eps}
\caption{As Fig. \ref{fig: JPG bivlf} but for the neural network morphological type. The trends in Fig. \ref{fig: JPG bivlf} become clearer due to the sample size of 20,891 compared to 998. \label{fig: ANN bivlf}}
\end{figure}

The LF bivariate with the JPG catalogue morphological type $\tjpg$ obtained here is shown in Fig. \ref{fig: JPG bivlf} and that with the ANN morphological type $\tann$ in Fig. \ref{fig: ANN bivlf}. The corresponding completeness maps (number of galaxies with spectra / number of galaxies, in each bin) are shown in Figs. \ref{fig: JPG completeness} and \ref{fig: ANN completeness}. There are no obvious biases within the regime for which galaxies are present, the completeness being consistent with 90\% for the full range of apparent magnitude for $\tjpg$ and 60\% for $\tann$, the latter being expected from the use of the whole survey area which is only about 70\% complete in spectroscopy (Appendix \ref{app: completeness maps}). Fig. \ref{fig: ANN completeness} again shows the lack of types later than 4.5 assigned by the ANN (see \S \ref{sec: nn types}). If the training set without the clean photometry flags were used, hence leaving in more late type galaxies, the $\psi$ greyscale would extend further right by approximately the width of one of the bins, so the LF for very late types would still not seen.

The bivariate luminosity-morphology distribution was computed for the same JPG catalogue and classification scheme by \citet{nakamura:morphlf}. However, whilst being drawn from the same catalogue there are some differences in the method, leading to an altered final sample for the LF. A different subset of the EDR galaxies is used here, with 1,150 instead of 1,482; the sample is divided into smaller bins and the normalisation is without the explicit redshift cut used in \citet{nakamura:morphlf}. The smaller value is 1,150 due to the masking applied to the datasets here. Of these, 1,019 have redshifts and 998 are selected for the LF.

\citeauthor{nakamura:morphlf} show that the overall LF is well fit by a Schechter function with parameters $\alpha = -1.10$, $M^* = -20.65$ and $\phi* = 0.0143$. We obtain $\alpha = -1.30$, $M^* = -20.58$ and $\phi* = 0.0098$. Their LF is divided into the ranges $0 < \tjpg < 1.5$, $1.5 < \tjpg < 3$, $3.5 < \tjpg < 5$ and $5.5 < \tjpg < 6$. For 0--1 and 3.5--5 the faint-end slope is shallower, at $\alpha = -0.83$ and $\alpha = -0.71$. 1.5--3 is steeper, at $\alpha = -1.15$. The early types are also brighter, as expected, being $M^* = -20.75$ compared to $M^* = -20.30$ for types later than 1. In our LFs there is no clear trend between the types, although the binning is narrower and the earliest type bin appears brighter and with a shallower faint-end slope as expected.

In the LF split by $M_r$, the main trend is for higher normalisation at fainter magnitudes, as expected for increased numbers of fainter galaxies. For bright galaxies, the normalisation with $\tjpg$ increases from $\phi \sim 10^{-4}$ to $\phi \sim 5 \times 10^{-3}$ between types 1 and 0 but otherwise is consistent with a flat distribution across the range in $\tjpg$ at all magnitudes, with a possible decrease at types later than $\tjpg = 4$.

The neural network types enable the LF sample to be enlarged from 998 to 20,891 galaxies. This is the first time the LF with morphology assigned in this way to the resolution of individual Hubble types has been calculated for a sample of galaxies of this size. The trends hinted at in Fig. \ref{fig: JPG bivlf} become clearer, with the types earlier than $\tann = 0.5$ now showing a declining faint-end slope. The $0.5 < \tann < 1.5$ are as bright but more abundant at fainter magnitudes. The spirals from $\tann = 1.5$ to $\tann = 3.5$, corresponding to S0a to Sbc, then show no difference in their LFs. The late types to 4.5 are then slightly fainter again, with a slightly steeper faint-end slope. Each slice appears consistent with a Schechter function. For the LF split by $M_r$, the galaxies brighter than $\mrlogh = -21$ are now clearly more abundant at $\tann < 1$, with a weaker trend seen for $-21 < \mrlogh < -20$. The rest to $\mrlogh = -18$ are still consistent with a flat distribution, except at $\tann < 0.5$ where they decline. The changes at the early type end are at least as strong as those indicated, as the ANN bias here is, as mentioned, away from early types and not towards. There are very few galaxies later than $\tann = 2.5$ and brighter than $\mrlogh = -22$.

In \citet{delapparent:morphlf}, it is suggested that cross-contamination between Hubble types will make the LFs tend towards Schechter functions even if the underlying LFs for the giant (as opposed to dwarf) galaxies are Gaussian. Here the $-0.5 < \tann < 0.5$ bin in particular could be suffering from this effect if the true distribution were Gaussian brighter than $\mrlogh = -20$ and the fainter magnitudes were contaminated by dwarfs. \citet{nakamura:morphlf} do not find this to be a significant effect in their results as their fainter galaxies do not have the softer cores of later type dwarfs. However, the Gaussian LF is supported by fig. 1 of \citet{bernardi:elliptical2}, who show LFs for ellipticals with a turnover at $\mrlogh \sim -20.5$. The LFs for the later types will similarly contain dwarf and giant galaxies and will therefore also appear Schechter-like.

Several recent papers discuss blue dwarf spheroidal galaxies. These have the same concentrations and smooth appearance as dwarf ellipticals, but are much bluer in colour. Whilst much more prevalent at higher redshifts \citep[e.g.][]{ilbert:morphlfevoln}, for local galaxies these objects are discussed by \citet{driver:mgcmorph}. In their fig. 6, they show that when the blue spheroids are removed, the E/S0 LF gains a much more rapidly declining faint-end slope, with $\alpha$ changing from -0.9 to -0.4. This could be addressed in our work by adding colour as a classification criterion, moving towards a spectro-morphological classification.

It may also be the case that the morphological types are not well defined below a resolution of about 0.5 T types and will therefore inevitably be cross-contaminated. However with the large sample here the LF differences are significant, as indicated by the sizes of the error bars.

Another recent determination of the LF bivariate with Hubble type is presented in \citet{read:massfn}. This shows LFs following the same trends as seen here, and with only those for types Sa and Sb being similar.

\subsection{Inverse Concentration Index} \label{sec: LF CI}

Various papers \citep[e.g.][]{shimasaku:brightgal} have shown that the inverse concentration index $R_{50}/R_{90}$ ($\ciinv$) is correlated with the morphological type, and indeed can be used as a simple classifier. Here we show the LF bivariate with concentration index in Petrosian radii, using the same $R_{50}/R_{90}$ ratio.

\citeauthor{shimasaku:brightgal} show that, whereas the $\ciinv$ does correlate with morphology, the correlation is not perfect. In particular, for types E and S0, the index varies from around 0.3 to 0.35, whereas it increases to almost 0.5 to type Sc. It then stays in this range for even later types.

\begin{figure}
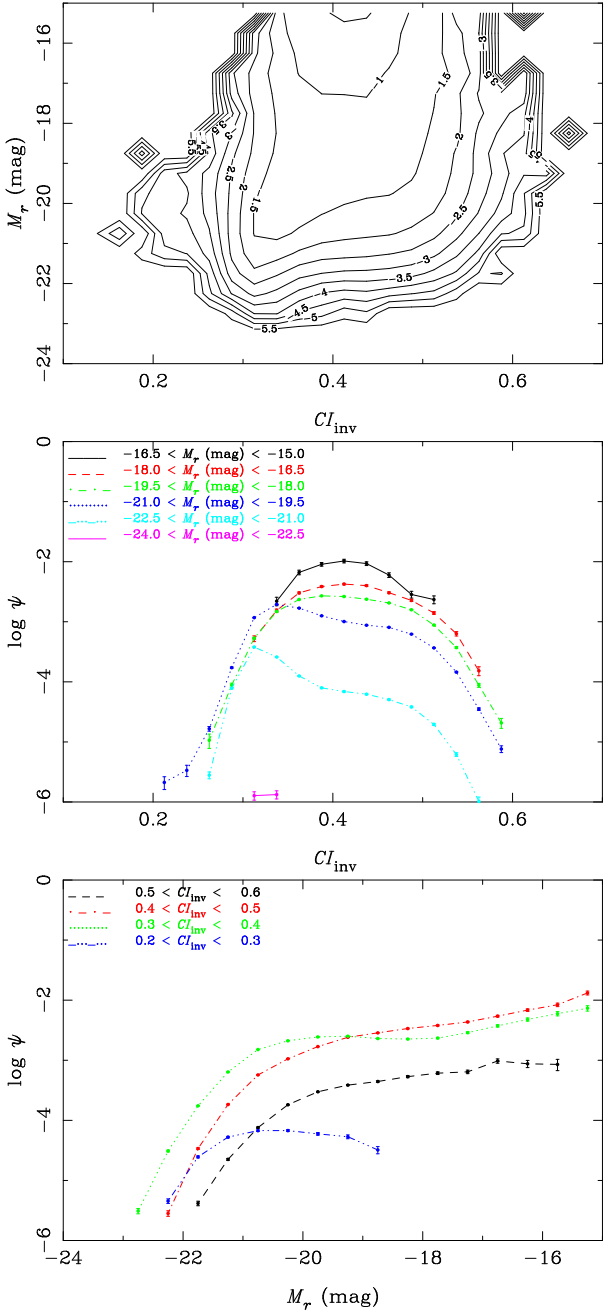
 \centering
\includegraphics[height=0.45\textwidth,angle=270]{figures/figure_03_panel1.eps}
\includegraphics[height=0.45\textwidth,angle=270]{figures/figure_03_panel2.eps}
\includegraphics[height=0.45\textwidth,angle=270]{figures/figure_03_panel3.eps}
\caption{LF bivariate with $R_{50}/R_{90}$ inverse concentration index. The expected trend of fainter galaxies being less concentrated is seen, but since the result is based on 251,744 galaxies (see Table \ref{table: bivlfs}), much further detail is visible (see text). \label{fig: invCI bivlf}} 
\end{figure}

The bivariate luminosity-$\ciinv$ distribution is shown in Fig. \ref{fig: invCI bivlf} and the completeness map in Fig. \ref{fig: invCI completeness}. The expected broad trend of fainter galaxies being less concentrated is seen. Subdivided by $\ciinv$ the LF resembles a Schechter function with a steeper faint-end slope for less concentrated galaxies. For the bins outside $0.3 < \ciinv < 0.6$ the normalisation is lowered due to lack of galaxies. The completeness similarly becomes very low and noisy outside this range. For $0.3 < \ciinv < 0.4$ the LF shows an upturn at magnitudes fainter than $\mrlogh = -18$. This may be related to the prevalence of dwarf ellipticals in clusters and is discussed further below.

Subdivided by $M_r$, at the bright end, $-24 < \mrlogh < -21$, the $\ciinv$ shows a peak at $\ciinv \approx 0.31$. A pure de Vaucouleurs profile corresponds to $\ciinv = 0.3$, therefore this peak corresponds to the population of ellipticals. The slight offset above 0.31 may be due to the bias towards the $\ciinv$ for a PSF, which is 0.5 (Gaussian), due to many of the galaxies in the sample being poorly resolved, in the sense that their isophotal axis ratio rather than their shape measure is used (\S \ref{sec: properties}). However, this region is also populated by S0 galaxies. At $-21 <\mrlogh < -19.5$, the peak at low $\ciinv$ becomes broader and less pronounced, increasing to $\ciinv \approx 0.33$. At $-19.5 < \mrlogh < -18$ the distribution becomes more symmetrical, peaking around $\ciinv \approx 0.38$. At $-18 < \mrlogh < -16.5$ the distribution is approximately symmetrical around $\ciinv = 0.43$, which corresponds to an exponential profile. Besides the bright peak, the distributions are not otherwise obviously bimodal, just broad.

The completeness increases from $\sim$ 40\% to 50\% over the range $0.3 < \ciinv < 0.4$ and remains at this level until $\ciinv \sim 0.55$. This may reflect the bias due to the closest possible spacing of 55 arcsec for SDSS spectroscopic fibres (\S \ref{sec: data}): more galaxies in environmentally dense regions, which tend to be early type and therefore of $\ciinv$ in this range, will be missed. If this is the case then the peak in low $\ciinv$ seen should be slightly more pronounced than shown.

\citet{blanton:edrlf} present the LF bivariate with the Petrosian $\ciinv$ for the SDSS EDR. The same broads trends as seen here are present: a peak near $\ciinv = 0.3$ corresponding to galaxies with a de Vaucouleurs profile, particularly evident for the brightest galaxies at $-23.5 < \mrlogh < -22$, then becoming less prominent as the distribution broadens to spread as far as $\ciinv \sim 0.5$ at $\-20.5 < \mrlogh < -19.0$ and becoming symmetrical about $\ciinv \sim 0.43$, the exponential profile, at $\mrlogh < -17.5$. The de Vaucouleurs peak is offset slightly above 0.3 in a similar way to our results. Unlike our LF, the normalisation split by $M_r$ increases then decreases, whereas ours increases monotonically from bright to faint, presumably due to a different method of normalisation. Subdivided by $\ciinv$, the LFs show the usual trend of bright and shallow faint-end to dim and steep faint-end. Compared to our results which cut in axis ratio, there is no obvious difference in the widths of the distributions in $\ciinv$.

\citet{nakamura:morphlf} also present the bivariate luminosity-$\ciinv$ distribution for their JPG sample described in \S \ref{sec: LF T}, dividing the sample at $\ciinv = 0.35$ into early and late types. As with our results above $\ciinv = 0.3$, the LF does not show a marked decline at the faint end. They show that their sample is not dominated at the faint end ($M_r \simgt -19$) by dwarf ellipticals, which have softer cores.

\subsection{S\'ersic Index} \label{sec: LF Sersic}

\begin{figure}
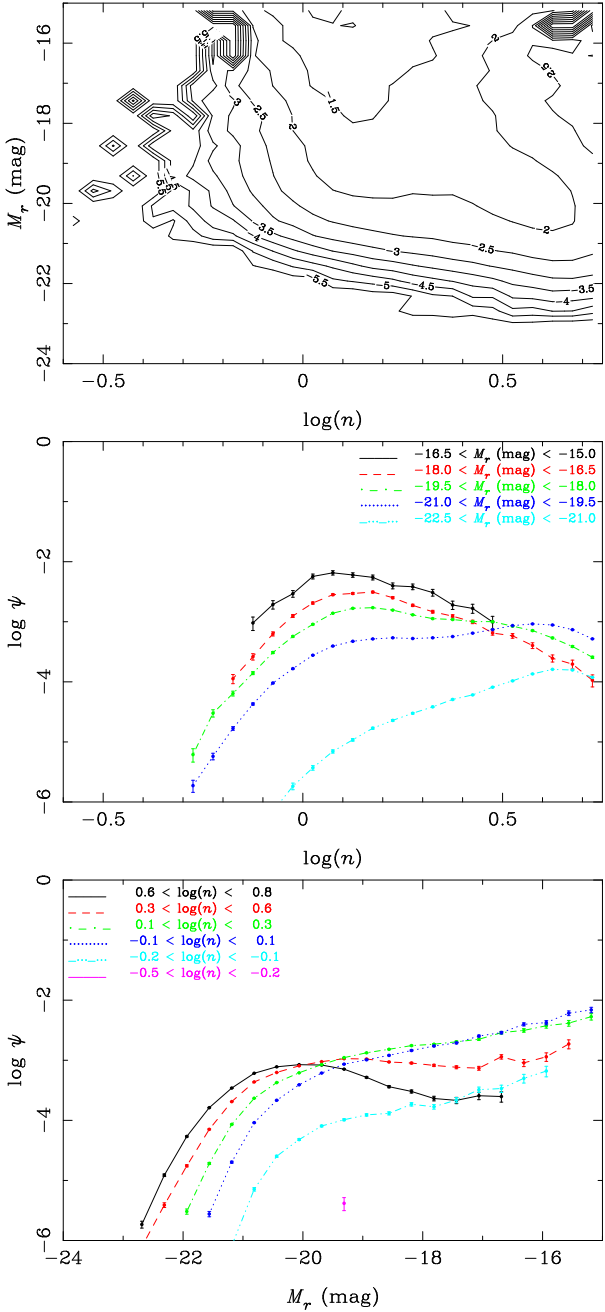
 \centering
\includegraphics[height=0.45\textwidth,angle=270]{figures/figure_04_panel1.eps}
\includegraphics[height=0.45\textwidth,angle=270]{figures/figure_04_panel2.eps}
\includegraphics[height=0.45\textwidth,angle=270]{figures/figure_04_panel3.eps}
\caption{LF bivariate with seeing-corrected S\'ersic index $n$. The Gaussian-like LF for the galaxies of high, $n$, corresponding to early morphological type, is clearly seen. \label{fig: Sersic bivlf}}
\end{figure}

The seeing-corrected S\'ersic index is monotonically related to the S\'ersic-radius-based concentration index \citep{graham:totalmag} and as such should show similar trends to those for the Petrosian-based $\ciinv$ presented in \S \ref{sec: LF CI}. However, unlike our concentration index, the S\'ersic index is seeing-corrected. We plot the index logarithmically as explained in \S \ref{sec: dr4}. The bivariate LF is shown in Fig. \ref{fig: Sersic bivlf} and the completeness map in Fig. \ref{fig: Sersic completeness}. The indices again correspond to profile types, with $\logn = 0$ ($n = 1$) being exponential and $\logn = 0.6$ ($n=4$) de Vaucouleurs. The completeness decreases from around 55\% at $\logn \sim -0.2$ to 40\% at $\logn = 0.65$. Again this could be due to bias against early types due to fibre spacing.

Subdivided by $M_r$, the distribution is clearly bimodal to fainter magnitudes than seen for $\ciinv$, down to $\mrlogh = -18$. The high-$n$ peak is around $\logn = 0.65$ for $-24 < \mrlogh < -21$, higher than de Vaucouleurs galaxies. For $-21 < \mrlogh < -19.5$ the peak shifts lower, to $\logn \sim 0.575$. Below $\mrlogh = -19.5$, the low-$n$ peak becomes prominent and is centred at $\logn \sim 0.15$. Again this is higher than the exponential profile of $\logn = 0$. At this value, the number of galaxies has dropped sharply, although the completeness remains flat until $\logn \sim -0.2$. 

Subdivided by $n$, the LFs are again Schechter-like and go from bright and shallow faint-end to dim and steep faint-end. As with $\ciinv$, the faint end of the LF for high $n$, whilst decreasing, does not resemble a Gaussian. Thus if contamination by dwarfs affects the faint-end LF for morphology and $\ciinv$, it is also at work for $n$. The LF turnup at $\mrlogh \simlt -18$ seen for $\ciinv$ is less prominent here, merely hinted at for $0.3 < \logn < 0.8$.

As described in \S \ref{sec: dr4}, the indices are systematically offset, so that a true de Vaucouleurs galaxy with $\logn = 0.6$ is here assigned $\logn \sim 0.55$. However, the value of $\logn$ for a PSF is -0.3 ($n = 0.5$), so this effect may again be at work, although the PSF is convolved in the S\'ersic model fitting so this seems unlikely.

The bivariate luminosity-$\logn$ distribution is also presented by \citet{driver:mgcmorph} for the MGC. There it is subdivided into $n \geq 2$ and $n < 2$ ($\logn = 0.3$) and fitted with a Schechter function. The respective parameters change from $\alpha = -1.25$ to $\alpha = -0.66$, $M^* = -19.48$ to $M^* = -19.35$ and $\phi* = 0.0129$ to $\phi* = 0.0087$. The high-$n$ LF again is Schechter-like. The less obvious distinction in their two distributions than those seen in their LFs for the MGC continuum measure and colour is ascribed at least in part to dwarf ellipticals, which spread across the $\logn = 0.3$ boundary. As with our results from \S \ref{sec: LF T}, colours are needed, which they go on to consider, finding a clear bimodality in the bivariate distribution of core $u-r$ colour and $n$. This and their spread in Sabc morphologies between the two peaks is argued as evidence for bulges and discs being the fundamental components and is discussed further in the context of our results below.

\subsection{Surface Brightness} \label{sec: LF SB}

The bivariate luminosity-surface brightness (SB) distribution has been investigated by several previous authors and is also known as the bivariate brightness distribution (BBD). It is important as it allows one to better quantify selection effects in a survey. Here we compare our results to those of \citet{blanton:edrlf}, \citet{shen:sdssgalsizes} and \citet{blanton:lflowl} from the SDSS and \citet{driver:mgcbbd} from the Millennium Galaxy Catalogue. Further recent studies at low redshift in the optical are \citet{dejong:sbsdmlf}, \citet{cross:ldens} and \citet{cross:bbd}. The former studies a more local sample of spirals in more detail and the latter two contain slightly earlier results than the MGC, from the 2dFGRS. The SDSS studies use similar data and the same passbands as our data. The MGC represents the current state-of-the-art in the optical by probing to a deeper limiting surface brightness of $26 \magarcsec$ in the $B$ band over smaller but still substantial ($37.5 \sqdeg$) area of sky. The broad trends here and in other recent work are in agreement with earlier work \citep[e.g.][etc.]{freeman:expprofile,phillipps:selection}, but are seen at higher signal to noise.

In the SDSS, the Petrosian and S\'ersic radii are derived from axisymmetric profiles. As neither the Petrosian \citep[\S \ref{sec: dr4},][]{blanton:dr1lf} nor the S\'ersic apertures are missing much flux for most galaxies, the SBs will therefore be systematically overestimated as the semimajor axis of a half-light elliptical aperture is always at least as large. The MGC study uses elliptical apertures.

\begin{figure}
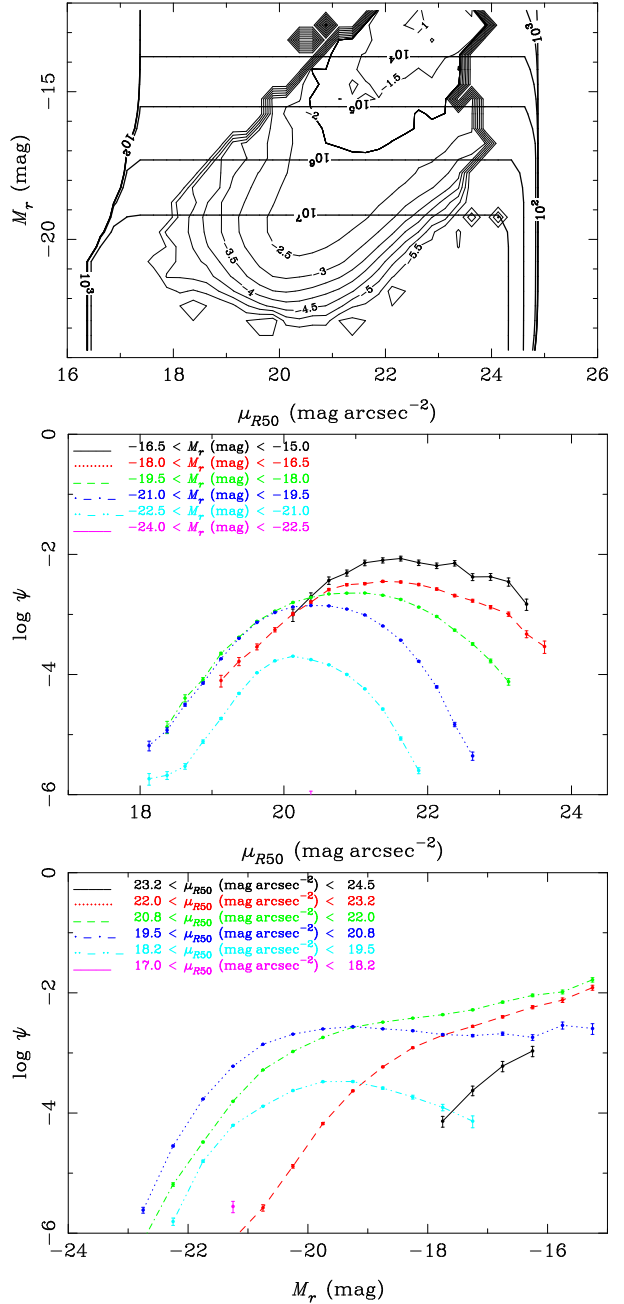
 \centering
\includegraphics[height=0.45\textwidth,angle=270]{figures/figure_05_panel1.eps}
\includegraphics[height=0.45\textwidth,angle=270]{figures/figure_05_panel2.eps}
\includegraphics[height=0.45\textwidth,angle=270]{figures/figure_05_panel3.eps}
\caption{LF bivariate with absolute effective Petrosian 50\% light radius surface brightness. The Petrosian radii are not seeing corrected. The distribution lacks bimodality and confirms the trend of fainter galaxies having lower surface brightness. The dataset contains an explicit surface brightness cut at $\sbr = 24.5$. The axis limits for the top panel are extended and the thick contours show the regions inside which volumes of $10^2$--$10^7 \cubicmpc$ were probed. \label{fig: SB bivlf}} 
\end{figure}

The bivariate LF is shown in Fig. \ref{fig: SB bivlf} and the completeness map for apparent surface brightness is shown in Fig. \ref{fig: SB completeness}. Also shown are contours at $10^2$--$10^7 \cubicmpc$ which show the regions inside which these volumes were probed by the survey. Note that the axes of the top panel are extended relative to the LF slices shown in the lower two panels. This shows that normalisation of LF drops before the volume becomes small due to limiting surface brightness and that the volume at $\mrlogh < -15$, for which the LF is quoted, is at least of order $10^5 \cubicmpc$. As a function of $\sbrapp$, the completeness varies between 45\% and 60\% for $19 \simlt \sbrapp \simlt 23$ and outside this range becomes noisy and drops to being consistent with zero at $\sbrapp \approx 17$ and $\sbrapp \approx 24$. As a function of apparent magnitude, the completeness is between 55\% and 60\% in the range $12 \simlt r \simlt 16.5$ and drops outside this to 40\% at $r = 17.77$, and low values at the bright end, consistent with zero brighter than $r \approx 11.5$. There is a clear lack of faint high surface brightness or bright low surface brightness galaxies. Fig. \ref{fig: SB completeness} shows the incompleteness for spectra and, as described in Appendix \ref{app: completeness maps}, \citet{blanton:lflowl} estimate the incompleteness for the photometry. Their ranges at which the completeness dropped below 90\% were $\sbrapp > 22$, when both $\sbrapp < 19$ and $r > 17$, and $\sbrapp \simgt 18.7$. They show that the SDSS galaxy distribution drops off well inside these completeness limits. Therefore, our distribution will similarly and the overall LF is unaffected away from the bright and faint end.

In the LF subdivided by $M_r$, we see the same overall distribution and trends as the previous studies. The $\sbr$ is a lognormal distribution with a roughly constant peak at bright magnitudes. This then broadens and dims at fainter magnitudes. The peaks are $\sbr \approx$ 20.1, 20.5, 21.0, 21.5 and $\sim 21.75$ in steps of 1.5 magnitudes for $-22.5 < \mrlogh < -15$. No bimodality is evident, meaning that if the galaxies are divided into one or more populations, as suggested by some of the other measures here such as colour, the surface brightness LF is insensitive to it, or the distributions overlap to such an extent that there is no central dip, with perhaps just a broadening.

Subdivided by $\sbr$, the LFs appear Schechter-like, with a fainter $M^*$ and steeper faint-end slope towards fainter surface brightness, as expected, although the brightest bins lack galaxies. This is also consistent with previous results.

\citet{blanton:edrlf} present the BBD for the SDSS EDR. Although the overall luminosity density found by them has been superseded by \citet{blanton:dr1lf}, their BBD is consistent with ours. Petrosian magnitudes are used. They also use the EDR calibrations of the passbands $u^*$, $g^*$, etc. but the difference here is small. They find for $-23.5 < M_{r^*} < -20.5$ that the peak is at $\sbr \sim 20$, the same as ours for $-22.5 < \mrlogh < -19.5$. The distribution then broadens and the peak dims to $\sbr \approx 20.5, 21.5$ and 22 over the steps of $1.5 \mgn$ to $\mrlogh < -16.0$. Given the slightly different binning and the broadness of the peaks these are consistent with our values. At the bright end ($M_{r^*} < -19$) the distribution drops off well above the EDR surface brightness cut of $\sbr < 23.5$ but at the faint end this may be an issue, as they are not able to show the upturn in the faint-end LF later reported by \citet{blanton:lflowl}. We see similar results relative to the $\sbr < 24.5$ cut. \citeauthor{blanton:edrlf} also plot the LF versus SB, finding as we do that the LF appears Schechter-like, with a fainter $M^*$ and steeper faint-end slope towards fainter surface brightness. We see a clearer downturn in the LF in the range $18.2 < \sbr < 19.5$.

\citet{shen:sdssgalsizes} use data from a sample similar to the SDSS DR1. In their fig. 14, they show the surface brightness distributions for early and late type galaxies, defined by concentration index less than and greater than 2.86 respectively. The measures use S\'ersic magnitudes and $h=0.7$, which we convert to $h=1$ here. Late types show mean SBs of around $20.2 \magarcsec$ for $-21.2 < \mrlogh < 20.2$. This then drops to $20.5 \magarcsec$ for $\mrlogh = -18.2$ and more rapidly to $21.5 \magarcsec$ for $\mrlogh = -15.2$, the faintest level. The early types are around $20 \magarcsec$ for $\mrlogh = -21.2$, $19.5 \magarcsec$ for $\mrlogh = -20.2$ to $19 \magarcsec$ for $\mrlogh = -18.2$, the faintest. The distributions again broaden at fainter magnitudes. These are approximately consistent with our results.

\citet{blanton:lflowl} use the SDSS DR2 and show the overall Petrosian SB distributions for $-20.5 < M_r < -12.5$. Over the range $-20.5 < M_r < -15$ their peak varies from $20.6 \magarcsec$ to $22.6 \magarcsec$. This compares to our $20.5 \magarcsec$ and around $21.75 \magarcsec$ over the same range.

\citet{driver:mgcbbd} study the BBD in the MGC, using an improved SB definition involving elliptical isophotes and the deeper $\sbrapp$ limit mentioned of $26 \magarcsec$ in $B$. Due to the overestimation in the SDSS mentioned above, they find a systematic offset of $0.4 \mgn$ to fainter values of the SB compared to \citet{shen:sdssgalsizes}. They also see the more constant SB values at bright $M_r$ but are otherwise consistent with \citet{blanton:lflowl}. 

An improvement to the study here would be to correct for the effects of internal extinction due to dust. This is particularly strong in highly inclined spiral discs. In this paper we attempted to reduce this effect by excluding galaxies with an axis ratio greater than an E7 elliptical using the axis ratio cut described in \S \ref{sec: properties}. A recent study which does apply a dust correction is \citet{dejong:sbsdmlf}. They use a more local sample of galaxies from \citet{mathewson:mfb} and \citet{mathewson:mfb2}, consisting of approximately 1,000 spirals of types Sb--Sdm. A detailed set of criteria that might be followed by future analysis are given by \citet{boyce:bbd}.

\subsection{Colours} \label{sec: LF Colours}

\begin{figure}
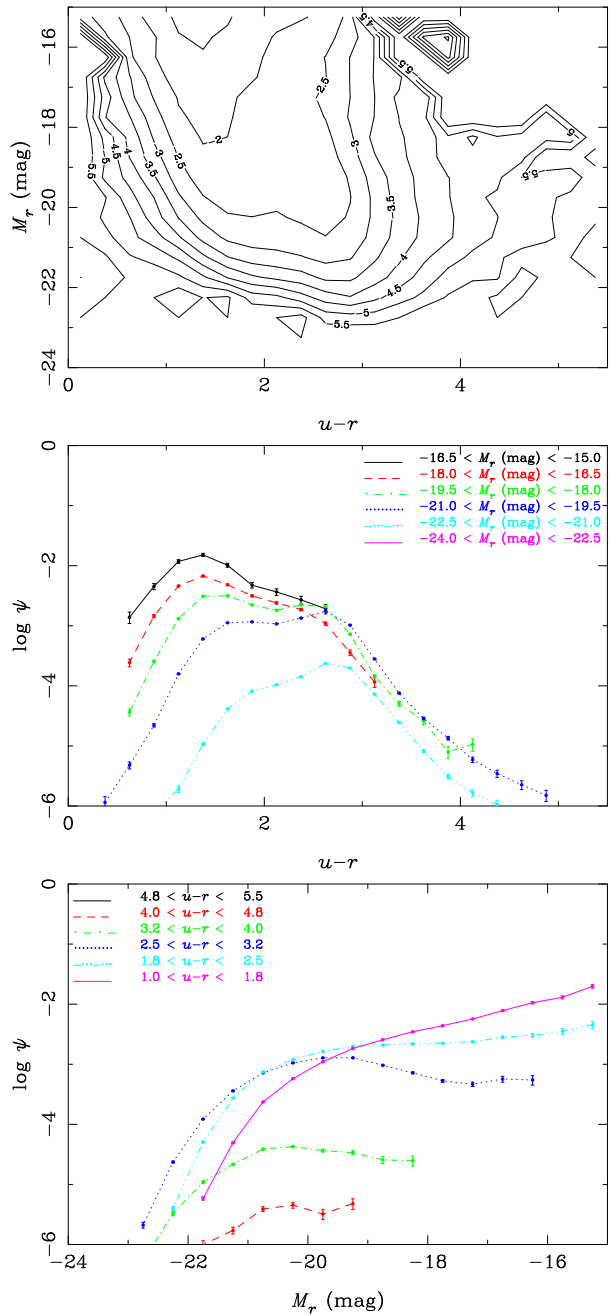
 \centering
\includegraphics[height=0.45\textwidth,angle=270]{figures/figure_06_panel1.eps}
\includegraphics[height=0.45\textwidth,angle=270]{figures/figure_06_panel2.eps}
\includegraphics[height=0.45\textwidth,angle=270]{figures/figure_06_panel3.eps}
\caption{LF bivariate with $u-r$ colour, extinction and K-corrected to $z=0.1$. The expected trends of a fainter $M^*$ and steeper $\alpha$ for bluer colours are seen. \label{fig: u-r bivlf}} \end{figure}

\begin{figure}
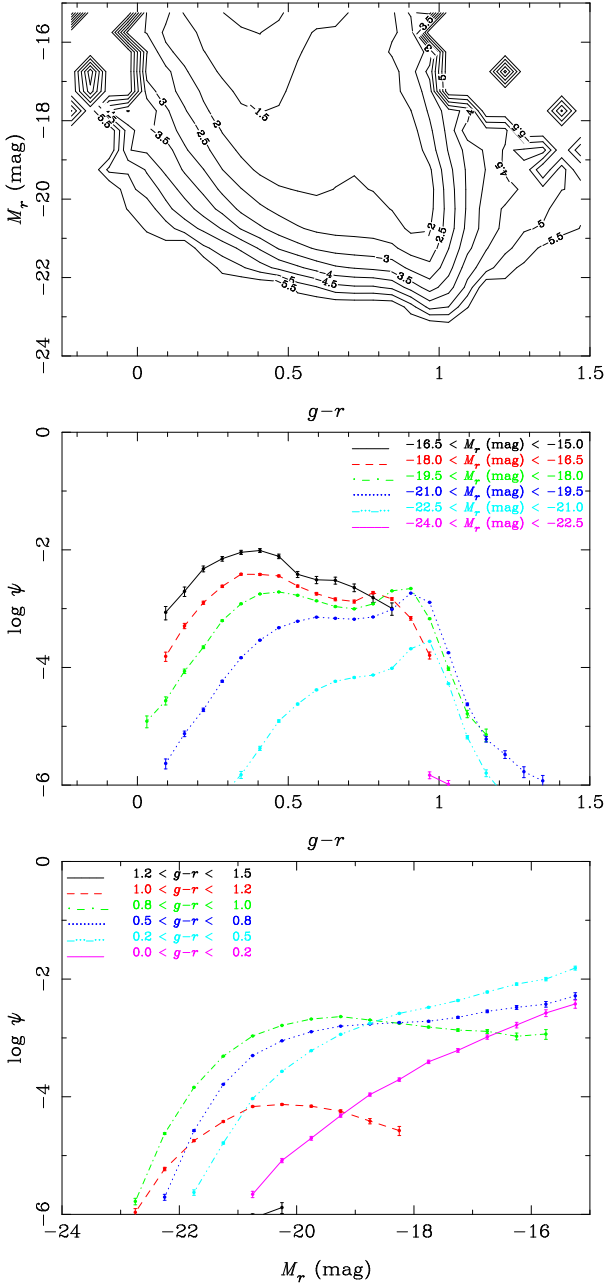
 \centering
\includegraphics[height=0.45\textwidth,angle=270]{figures/figure_07_panel1.eps}
\includegraphics[height=0.45\textwidth,angle=270]{figures/figure_07_panel2.eps}
\includegraphics[height=0.45\textwidth,angle=270]{figures/figure_07_panel3.eps}
\caption{As Fig. \ref{fig: u-r bivlf} but for $g-r$ colour. The results are qualitatively similar to $u-r$ with the red galaxy peak gaining strength. \label{fig: g-r bivlf}} 
\end{figure}

\begin{figure}
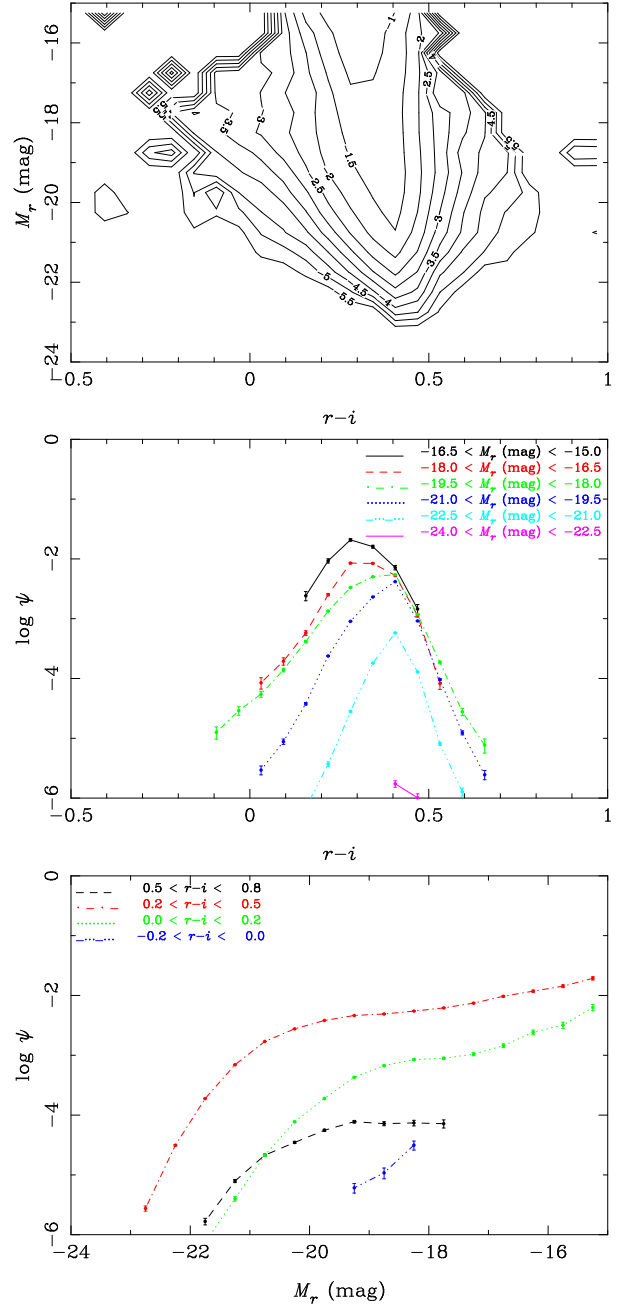
 \centering
\includegraphics[height=0.45\textwidth,angle=270]{figures/figure_08_panel1.eps}
\includegraphics[height=0.45\textwidth,angle=270]{figures/figure_08_panel2.eps}
\includegraphics[height=0.45\textwidth,angle=270]{figures/figure_08_panel3.eps}
\caption{As Fig. \ref{fig: u-r bivlf} but for $r-i$ colour. The blue of star formation is less pronounced. \label{fig: r-i bivlf}} 
\end{figure}

\begin{figure}
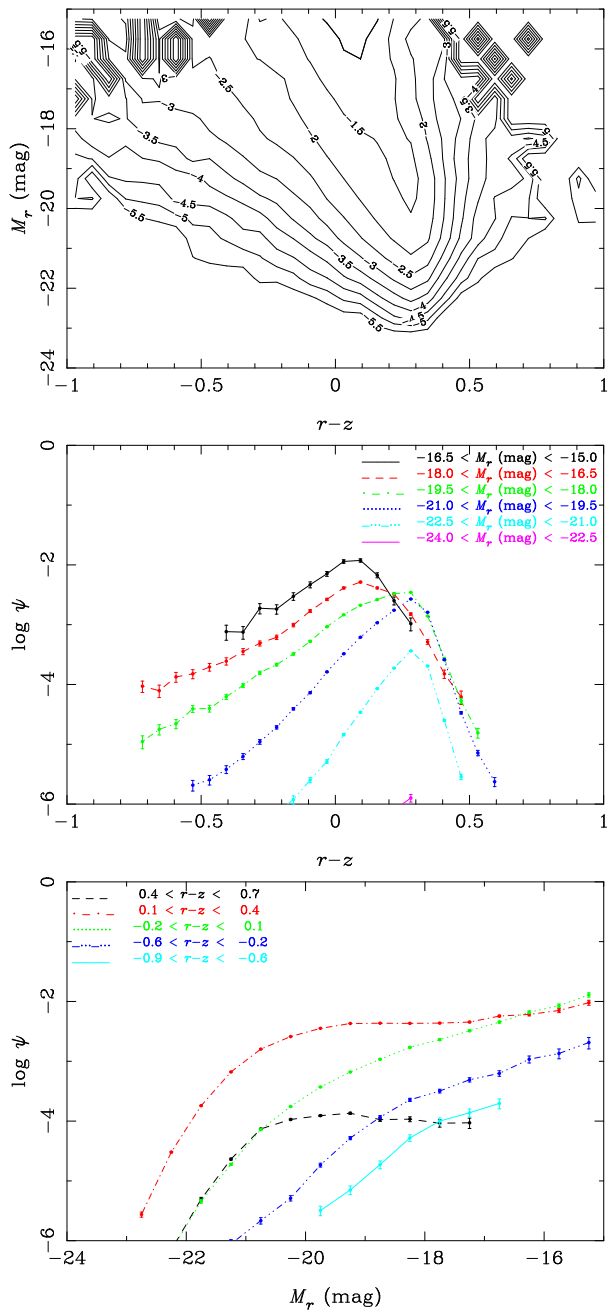
 \centering
\includegraphics[height=0.45\textwidth,angle=270]{figures/figure_09_panel1.eps}
\includegraphics[height=0.45\textwidth,angle=270]{figures/figure_09_panel2.eps}
\includegraphics[height=0.45\textwidth,angle=270]{figures/figure_09_panel3.eps}
\caption{As Fig. \ref{fig: u-r bivlf} but for $r-z$ colour. \label{fig: r-z bivlf}} 
\end{figure}

The colours are computed using the S\'ersic magnitudes K-corrected to $z=0.1$, the approximate mean of the sample. The $r$ band shifted to $z=0.1$ has $\lambda_{\mrm{eff}} \sim 5600$ \AA, similar to the V band. They are preferable to the observed frame colours as the galaxies over the whole redshift range are then more directly comparable. The LF fit implicitly assumes no evolution in the sample. One could take this into account by dividing the sample into redshift bins. All ten colours from $ugriz$ were investigated and Figs. \ref{fig: u-r bivlf}--\ref{fig: r-z bivlf} show the LF bivariate with $u-r$, $g-r$, $r-i$ and $r-z$, a representative sample of the results.

For the colours, the completeness (not shown) varies between 60\% and 70\% for $13 \simlt r \simlt 16$, dropping to 50\% by $r = 17.77$ at the faint end and to consistent with zero at $r \sim 11.5$.

In the LF split by colour, the expected trends of a fainter $M^*$ and steeper $\alpha$ for bluer colours are seen.  There is a clear lack of faint galaxies with redder colours in all four plots. The panels split by colour show that the LFs are Schechter-like, with a steepening faint-end slope with bluer colour and possible upturns fainter than $\mrlogh = -18$, apart from in the bluest bins where the slope is already steep.

In $u-r$ we see a constant peak around $u-r = 1.5$ for $-18 < \mrlogh < -15$. This then brightens to $u-r \approx 1.75$ by $\mrlogh = -21$ and disappears at brighter magnitudes. A red peak around $u-r = 2.5$ becomes evident at about the same magnitude and brightens to $u-r \sim 2.75$ by $\mrlogh = -22.5$.

The bimodality is clearer in $g-r$, with the blue peak showing similar behaviour to that in $u-r$, moving from $g-r \approx 0.35$ to $g-r \approx 0.6$ from $\mrlogh = -15$ to $\mrlogh = -21$. The red peak is visible much fainter than in $u-r$, moving from $g-r \approx 0.8$ through 0.9 to $g-r \sim 1$ by $\mrlogh = -24$.

In $r-i$, bimodality is no longer evident, with a single peak monotonically moving from $r-i \approx 0.3$ to $r-i \approx 0.4$ from $\mrlogh = -15$ to $\mrlogh = -21$. The peak is then approximately constant at $r-i \sim 0.4$ at brighter magnitudes. There is therefore either no bimodality in this colour, or the peaks due to the populations causing the peaks in $u-r$ and $g-r$ are indistinguishable, either intrinsically or at the colour resolution probed. The constant colour brighter than $\mrlogh = -19.5$ suggests that this could still correspond to the early type bulge-dominated population. The lack of an obvious blue peak is consistent with those peaks from $u-r$ and $g-r$ being due to star formation.

$r-z$ shows similar behaviour to $r-i$, with the peak moving from $r-z \approx 0.1$ to $r-z \approx 0.3$, again constant brighter than $\mrlogh = -19.5$.

The plots for $u-g$, $u-i$ and $u-z$ are qualitatively similar to $u-r$, those for $g-i$ and $g-z$ to $g-r$ and $i-z$ to $r-z$, which suggests no obvious physical process apparent in these colours which has been missed by the four shown.

The bivariate luminosity-colour distribution is also presented by \citet{blanton:edrlf} and \citet{driver:mgcmorph}. \citeauthor{blanton:edrlf} subdivide the SDSS EDR by Petrosian $g^* - r^*$ and see red and blue peaks at $g^* - r^* \sim 0.3$ and $g^* - r^* \sim 0.75$ respectively, with the same trends in the LF by colour. Our blue peak is the same but the red peak is at $g-r \sim 0.9$ in both Petrosian and S\'ersic magnitudes.

\citet{driver:mgcmorph} show the $u-r$ colour, divided at $u-r = 2.1$. They show the Petrosian value and that from the DR1 PSF magnitudes, which are superior at separating the bimodal population. For the core colours the Schechter parameters are $M^* = -19.15, -19.25$, $\phi^* = 0.0111, 0.0136$ and $\alpha = -0.15, -1.28$ for the two populations respectively, thus showing a clear downward-turning but not Gaussian LF for the red galaxies.

The prominence of the bluer population in $u$ and $g$ suggests that this peak in the LF is associated with star formation. This is supported by the GALEX LFs of \citet{wyder:galexuvlf}, \citet{treyer:galexlf} and \citet{budavari:galexlf}.

The $u-r$ colour is used in other investigations such as \citet{balogh:bimodallfenvt} and \citet{baldry:bimodalcmd} where the galaxies are clearly divided into two populations using counts in colour and morphology bins, which are fit by Gaussians. This is also done in B06 where $u-r$ is compared to the neural network morphological type.

Overall, the LFs subdivided by colour are consistent with a bimodal population of red bulges and star forming discs.

\subsection{Absolute Petrosian 90\% Radius} \label{sec: LF R90}

\begin{figure}
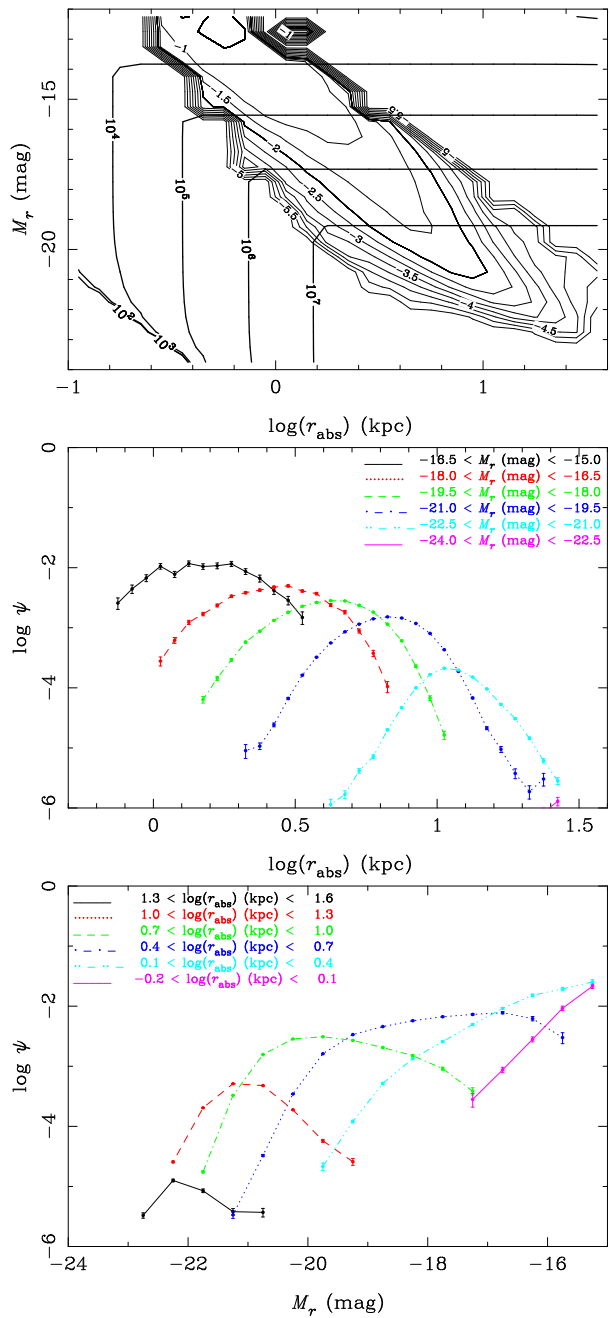
 \centering
\includegraphics[height=0.45\textwidth,angle=270]{figures/figure_10_panel1.eps}
\includegraphics[height=0.45\textwidth,angle=270]{figures/figure_10_panel2.eps}
\includegraphics[height=0.45\textwidth,angle=270]{figures/figure_10_panel3.eps}
\caption{LF bivariate with absolute Petrosian 90\% radius in kpc. The LF has a steeply declining faint-end slope at large galaxy sizes, turning into a rising faint-end slope at small sizes, suggesting that the steeper Schechter faint-end slopes seen in other plots are due to the small galaxies in the respective samples. As in Fig. \ref{fig: SB bivlf}, the axis limits in the top panel are extended and the thick contours show the regions inside which volumes of $10^2$--$10^7 \cubicmpc$ were probed. \label{fig: R90 bivlf}}
\end{figure}

The bivariate LF is shown in Fig. \ref{fig: R90 bivlf}. The completeness is 50--60\% for $13 \simlt r \simlt 16$, zero at $r \approx 11$ and 40\% at $r = 17.77$.

The expected pattern of fainter magnitude and higher normalisation due to larger numbers of galaxies with decreasing radius is seen. There is a clear cutoff with magnitude for a given radius, giving a declining faint-end slope at large galaxy sizes, which turns into a rising faint-end slope at small sizes. This suggests that the rising faint-ends in the Schechter LFs seen in other plots here are due to the smaller galaxies in a sample. As with surface brightness, the regions in which very small volumes are probed do not cut off the LF contours.

\subsection{eClass Spectral Type} \label{sec: LF eClass}

\begin{figure}
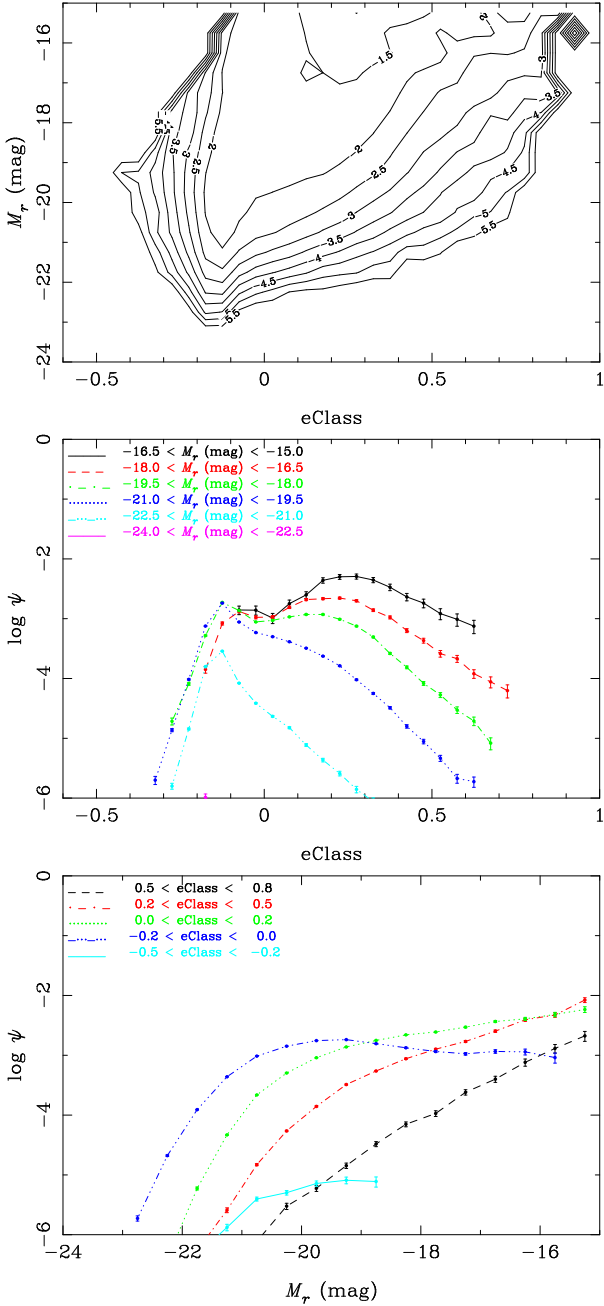
 \centering
\includegraphics[height=0.45\textwidth,angle=270]{figures/figure_11_panel1.eps}
\includegraphics[height=0.45\textwidth,angle=270]{figures/figure_11_panel2.eps}
\includegraphics[height=0.45\textwidth,angle=270]{figures/figure_11_panel3.eps}
\caption{LF bivariate with eClass PCA spectral type. There is a clear population of bright galaxies with low eClass which probably correspond to bright ellipticals. Some bimodality is evident. \label{fig: eClass bivlf} } 
\end{figure}

Fig. \ref{fig: eClass bivlf} shows the bivariate luminosity-eClass distribution. The eClass is not measured in the absence of a spectrum so there is no completeness map (although \citet{ball:ann} show that the eClass can be predicted to a RMS accuracy of $\pm 0.06$ using ANNs). The range $-0.6 \simlt \eclass \simlt 1.0$ corresponds to early--late type galaxies.

Subdivided by eClass, the LF shows the expected bright--shallow, dim--steep faint-end slope corresponding to early--late type galaxies. For $-0.2 < \eclass < 0.0$ the faint end LF is decreasing but not towards being consistent with a Gaussian LF.

Subdivided by $M_r$, there is a clear peak at $\eclass \approx -0.15$ for galaxies brighter than $\mrlogh = -18$ which becomes less distinct at fainter magnitudes. There is a second much broader peak for $\mrlogh = -19.5$ and fainter, moving from $\eclass \approx 0.15$ to $\eclass \approx 0.25$.

The two peaks are thought, as with the colours above, to correspond to the passive bulges of bright ellipticals and star-forming discs.

The LF has not been presented bivariate with eClass previously in the literature. However, two recent studies use the similar PCA-based 2dFGRS $\eta$ class. This covers the range $-5 \simlt \eta \simlt 10$, with low--high values corresponding to early--late types, in the same way as the eClass. \citet{madgwick:speclf} show LFs for $\eta < -1.4$, $-1.4 \leq \eta < 1.1$, $1.1 \leq \eta < 3.5$ and $\eta \geq 3.5$. For the first bin, the LF shows a pronounced increase in normalisation fainter than $\mrlogh = -16$, and appears Schechter-like in the other bins, with the usual dimming and steepening of the faint-end slope, although the Schechter function is always a poor fit. The $\eta < -1.4$ class is better fit by a Schechter + power law function.

\citet{driver:mgcmorph} shows the LF in the MGC for the classes $\eta = 1$, $\eta = 2$ and $\eta = 3,4$, showing a similar pattern to \citeauthor{madgwick:speclf} but without the faint-end upturns as the LFs are shown to $\mrlogh < -16$. Their MGC continuum type separates their LFs more clearly.

\citet{kochanek:inconsistencies} discusses possible biases in spectral LFs due to the use of spectroscopic fibres for the latter. The spectroscopic fibres will result in earlier spectral types for large galaxies due to bias toward their central regions. This bias is not quantified here, but their fig. 4 suggests that it is likely to be present even though most of our galaxies are poorly resolved (\S \ref{sec: properties}). However, it is unlikely to change the broad behaviour of the LF split by $M_r$, i.e. a roughly constant low eClass peak and a higher peak which increases in value for fainter galaxies. Also, various studies \citep[e.g.][]{glazebrook:cosmicspectrum} find that the effect is not large for star formation rate, which is a dominant constituent of the eClass and the $\eta$ type via emission lines.

\subsection{Other measures} \label{sec: LF Other Measures}

\begin{figure}
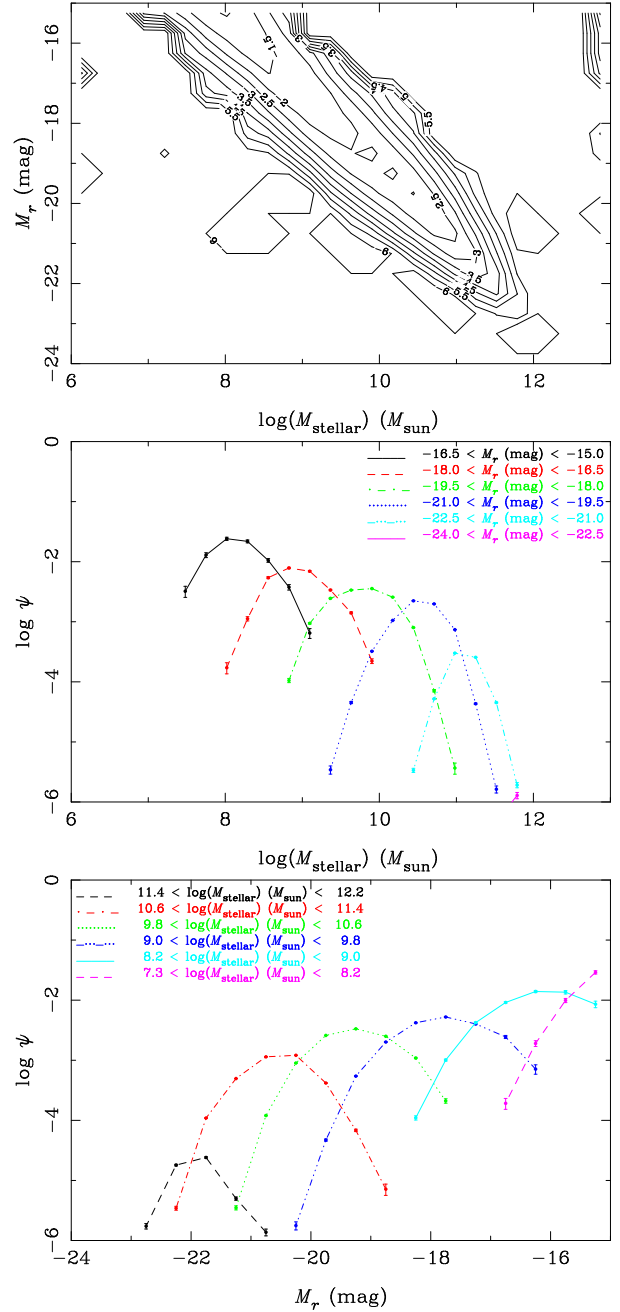
 \centering
\includegraphics[height=0.45\textwidth,angle=270]{figures/figure_12_panel1.eps}
\includegraphics[height=0.45\textwidth,angle=270]{figures/figure_12_panel2.eps}
\includegraphics[height=0.45\textwidth,angle=270]{figures/figure_12_panel3.eps}
\caption{LF bivariate with galaxy stellar mass. Although the $z$ band absolute magnitude is likely to correlate with stellar mass slightly more tightly than $r$, the $r$ band is shown for consistency with the other LFs. The correlation is expected as the galaxy luminosity is used in the estimation of stellar mass. \label{fig: Stellar mass bivlf}}
\end{figure}

\begin{figure}
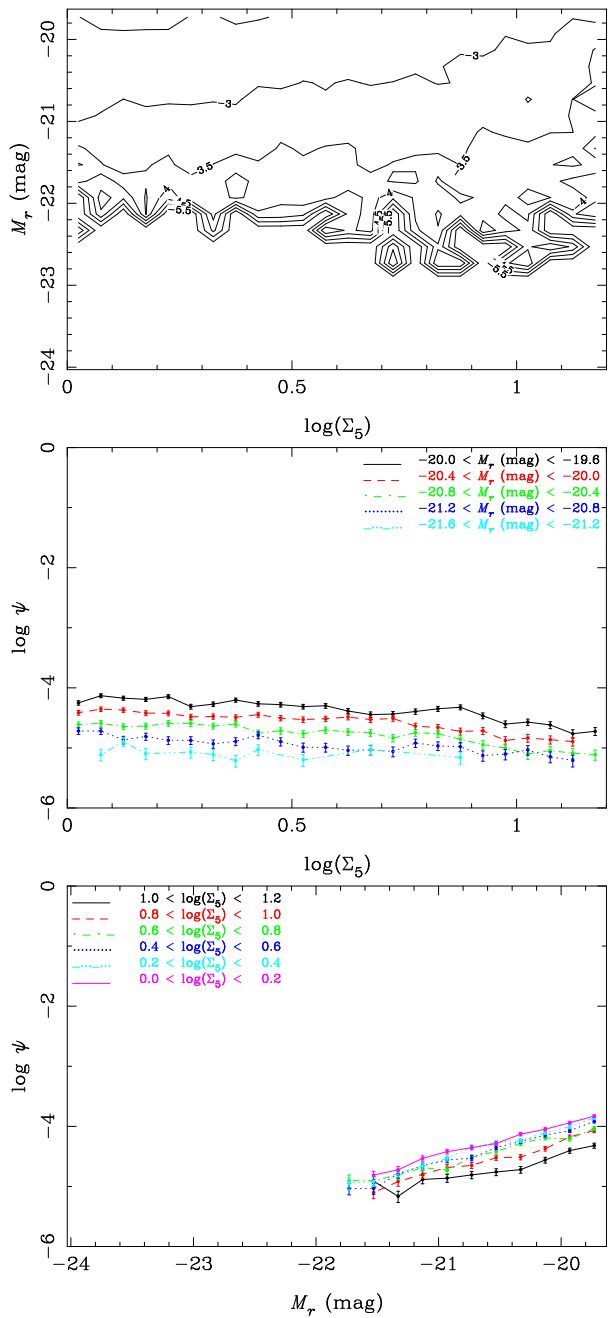
 \centering 
\includegraphics[height=0.45\textwidth,angle=270]{figures/figure_13_panel1.eps}
\includegraphics[height=0.45\textwidth,angle=270]{figures/figure_13_panel2.eps}
\includegraphics[height=0.45\textwidth,angle=270]{figures/figure_13_panel3.eps}
\caption{LF bivariate with $\Sigma_5$ surface density of galaxies. Unlike the other bivariate LFs here, the normalisation is arbitrary as it depends on the binning of the galaxies in density. \label{fig: Density bivlf}}
\end{figure}

Two further parameters were investigated: the galaxy stellar mass and the $\Sigma_N$ surface density of galaxies. These are shown in Figs. \ref{fig: Stellar mass bivlf} and \ref{fig: Density bivlf}, again with no completeness maps as these measures require spectra.

The stellar mass is useful as a physical parameter which could be directly compared with simulations. A tight correlation is seen between the LF and the stellar mass, with the LF in each bin showing a symmetrical shape of similar Gaussian form to that in absolute effective surface brightness. However, these results are expected because the luminosity is used in the calculation of the stellar mass in the first place, as part of the mass to light ratio calculated from the models of star formation history \citep[see][]{kauffmann:masssfh}. For consistency with the other LFs, the $r$ band is shown, although the correlation would be expected to be slightly tighter in $i$ and $z$, due to these being less dominated by short-lived bright stars than the bluer bands. In the LF subdivided by stellar mass, the galaxies with a stellar mass of $\mstellar < 10^9 \msun$ have distributions that are clearly truncated by the faint cut-off in absolute magnitude. This is evidence that there are further less massive galaxies not seen in this study.

The $\Sigma_N$ surface density of galaxies from the VAC is used in the morphology-density relation in B06. The bivariate LF appears to be of very similar form across all surface densities, with a power-law form, differing only in normalisation, which increases for both fainter magnitudes and lower densities. The faint luminosity cutoff due to the $z > 0.053$ limit on the density in the VAC is $\mrlogh = -19.63$, the top edge of the figure. The $z < 0.093$ limit does not have a strong effect on the bright luminosity galaxies.

The variation in the LF with density and galaxy spectral type is investigated by \citet{croton:lfenvt}, who find that the LF is fitted by a Schechter function over all densities, that the $\phi^*$ and $M^*$ vary smoothly over the density range but that the faint-end slope does not change much. Their normalisation takes into account the volume in which galaxies are present in the density bin used. Our normalisation here is just the overall value for the bivariate LF, so the relative normalisations of the bins are essentially arbitrary. This is also seen in the LF divided by $M_r$, in which one expects fewer galaxies at low density with normalisation weighted by volume.

It is also possible that the broad lack of change in the shape of the LF with density is related to the similar lack of change in slope in the correlation function divided by luminosity seen by e.g. \citet{connolly:edrcf} and \cite{zehavi:edrclustering}, but this would require further investigation, especially now that \citet{zehavi:gcfpowerlaw} describes departures from the power law.

\section{Function Fitting to the Bivariate LF} \label{sec: LF Fitting}

The Schechter function (equation \ref{eq: Schechter}) can, by taking $L/L^* = 10^{0.4(M^*-M)}$, be written as \begin{multline} \phi(M)~dM = 0.4~\mrm{ln}(10)~\phi^{*} \\ \times \mrm{exp}~[-10^{0.4(M^*-M)}][-10^{0.4(M^*-M)}]^{\alpha + 1}~dM . \end{multline} If one then applies Bayes' theorem to the bivariate function then it can be written as \begin{equation} \label{eq: Bayes} \psi(M,X)~dM~dX = \phi(M)~\varphi(X|M)~dM~dX , \end{equation} which gives two-dimensional functions to fit to the SWML estimate. In general any well-behaved function can be fitted, expressed using either side of equation \ref{eq: Bayes}.

One choice is to use a Schechter function for $\phi(M)$ and a Gaussian with mean $A(M-M^*) + B$ and standard deviation $\sigma_X$, where $A$ and $B$ are constants for $\varphi(X|M)$. This was first used by \cite{choloniewski:bivlf} and is known as the Cho{\l}oniewski function: \begin{multline} \varphi(X|M)~dM~dX = \frac{1}{\sqrt{2\pi\sigma_{X}^{2}}} \\ \times \exp\left[\frac{-(X - A(M-M^*) - B)^2}{2\sigma_{X}^{2}}\right]~dM~dX . \end{multline} This gives a 6-parameter fit. \citet{choloniewski:bivlf} actually used $M$ rather than $M-M^*$ but the latter scales better with our data.

Due to this greater dimensionality than the monovariate LF, a simple maximum likelihood fit is impractical, as the parameter space to explore is large. The method adopted here is the simplex search, which, while not of the greatest accuracy near the minimum, is robust to starting conditions and finds the minimum rapidly.

Here the Matlab implementation, known as fminsearch, is used. It implements the Nelder-Mead simplex algorithm \citep{nelder:simplex} using the method of \citet{lagarias:simplex} as described at http://\-www.\-mathworks.\-com. Options used were the Matlab default tolerance of $10^{-4}$ and a maximum number of iterations of 10000. The search can sometimes become stuck, for example, if the simplex decays to a lower number of dimensions due to the new point being too close to the centroid. However if the search converged it did so in less than this number of iterations.

The measure of goodness of fit, the error function, is the standard $\chi^2$, given by \begin{equation} \chi^2 = \left(\frac{\psi_{obs} - \psi_{fit}}{\overline \psi_{err}}\right)^2 . \end{equation} The fit is done to $\mrm{log}_{10} \psi$ rather than $\psi$ so the error is estimated using \begin{equation} \overline \psi_{err} = \frac{1}{2} \left[ \mrm{log}_{10} ( \psi_{obs} + \psi_{err}) - \mrm{log}_{10} ( \psi_{obs} - \psi_{err}) \right] , \end{equation} with $\psi_{err}$ from the information matrix determined as described above.

\section{Function Fitting Results} \label{sec: LF Fitting Results}

\begin{figure} \centering
\includegraphics[width=0.45\textwidth]{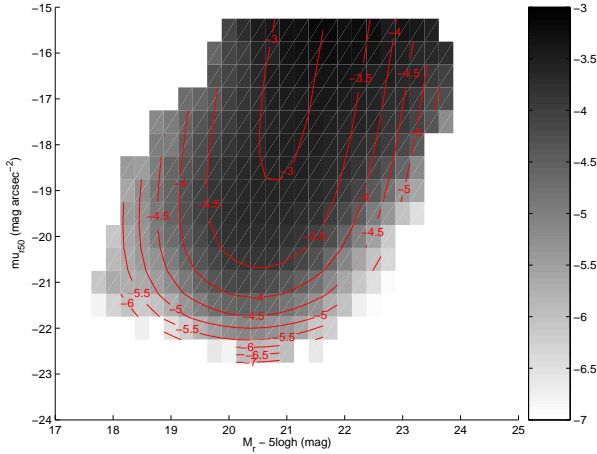}
\caption{SWML estimate in greyscale and fitted Cho{\l}oniewski function as contours for absolute effective Petrosian surface brightness. The right hand panel gives the LF $\mrm{log}_{10}~\psi$ value represented by the greyscale. The contours were fitted over the bins which contained five or more galaxies to minimize the effects of shot noise. The function provides a poor fit for bright galaxies. \label{fig: S-G bivlf fit}}
\end{figure}

\begin{figure} \centering
\includegraphics[width=0.45\textwidth]{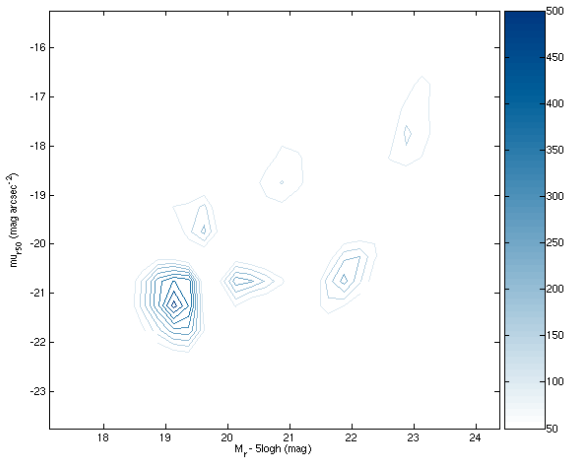}
\caption{$\chi^2$ values showing that the fit is generally poor, particularly for galaxies around $\mrlogh = -22$ and $\sbr = 20$. The right hand panel shows the $\chi^2$ values represented by the contours. \label{fig: S-G bivlf chisq}}
\end{figure}

It was generally found that the functions are a poor fit to the data, as the deviation of the data from a smooth function is large compared to the size of the error bars, giving a large $\chi^2$. Fig. \ref{fig: S-G bivlf fit} shows the fit of the Cho{\l}oniewski function to the BBD and Fig. \ref{fig: S-G bivlf chisq} the $\chi^2$ values. The fit is particularly poor for bright galaxies around $\mrlogh \sim -21.25$ and $\sbr \sim 19.25$ as the distribution narrows. The best-fitting parameters are $\phi^{*} = 0.0018, M^* = -20.6, \alpha = -1.19, \sigma_{\sbr} = 0.71, A = 0.30, B = 26.5$.

\citet{driver:mgcbbd} also find a poor fit, but for faint galaxies. They suggest that this is due to the assumption of a constant width for the Gaussian in absolute effective surface brightness whereas in fact the distribution broadens at fainter values. The assumption of a linear relation between the mean of the Gaussian and $M$ may also be a factor. It is unlikely that the simplex is always getting stuck in a local minimum far from the global minimum as the fit is poor from a variety of starting points but the minimum reached is the same. One could add further parameters but six is already a fair number. The results here are consistent with the constant Gaussian width approximation also being the problem. We confirm the suggestion of \citet{driver:mgcbbd} that bulge to disc decomposition is needed.

It is possible that no simple function will provide a good fit, as the superposition of underlying processes generating the bivariate LF may be too complex at the level of precision now available from the data. Or it may be the case that a single function will not fit, but the sum of two or more, each of which describes a different population, for example the red and blue or bulge and disc indicated above, may provide a better fit. The population might be subdivided using an optimal colour-separation criterion, for example that used by \citet{baldry:bimodalcmd}, then fitting, for example, a Cho{\l}oniewski function to each half. The difficulty is that the two distributions overlap, which would distort the two halves, and the sum of two Cho{\l}oniewski functions involves a large number of parameters, especially if the relation is non-linear between the mean of the Gaussian and $M$, the Gaussians vary in width, or skew-Gaussians are required.

Other bivariate LFs such as that with eClass were investigated with Schechter-Gaussian and Gaussian-Gaussian for earlier SDSS dat releases but the fit was also found to be poor. A skew-Gaussian was also tried using an available Matlab port of the software described at http://\-azzalini.\-stat.\-unipd.\-it/\-SN. The skew-Gaussian is given by \begin{equation} f(x) = 2\phi(x)\Phi(\alpha x), \end{equation} where $\phi(x)$ is the ordinary Gaussian and $\Phi(\alpha x)$ adds the skewness $\alpha$ \begin{equation} \phi(x) = \mrm{exp}~(-x^2/2)/\sqrt{2\pi} \qquad \Phi(\alpha x)  = \int_{-\infty}^{\alpha x} \phi(t) dt . \end{equation} A skewness of $\alpha = 0$ corresponds to the ordinary Gaussian, and a negative $\alpha$ gives the same shape as the positive but reflected about the vertical at $x=0$. The fits obtained were very similar to those for the ordinary Gaussian, with the skewness tending to low values of order 0.01.

\section{Discussion} \label{sec: discussion}

A major question raised by these results and other bivariate LF studies is whether a theory or simulation can be constructed which will correctly predict the distributions over the different second LF parameters presented. Two major themes in the LFs are the bimodality of the galaxy population and whether the underlying LFs per galaxy type are Schechter, Gaussian or some other form.

Here we fitted a Cho{\l}oniewski function to the bivariate luminosity-surface brightness distribution, and for earlier SDSS data releases experimented with it on other bivariate distributions and fitted Schechter, Gaussian, dual power-law Schechter and a sum of two Schechters to some of the individual LF slices. In all cases, the formal fits were found to be poor, with $\chi^2/\nu$ values of 10 or more. It is thus difficult to tell when a fit is the best that one can do and quote the numbers or if the fit really is poor and should be improved. We therefore have not quoted function fit parameters for individual LF slices in this paper, although with more work one could attempt to do so.

Concerning a bimodal population, the bivariate distributions are consistent with an early type, bright, concentrated, red population and a late type, faint, less concentrated, blue, star forming population. This idea of bimodality has been explored by others as it suggests two major underlying physical processes. These processes are thought to be connected to the formation of the bulge of the galaxy via mergers and the disc by accretion. Thus simulations of galaxy formation could be compared to the bivariate LFs presented. This is discussed further below.

Several results support the idea that the underlying LFs of the giant galaxies, i.e. the usual E, S0 and spirals are Gaussian, whereas those for the dwarfs are Schechter. \citet*{sandage:virgolf} and \citet{jerjen:morphlf} study the cores of nearby clusters; \citet{delapparent:morphlf} reviews the LFs subdivided by morphology in various surveys and \citet{delapparent:speclf} focusses on the ESO-Sculptor survey. A physical model supporting the idea is given in \citet{schaeffer:bimodallf} in which the maximum galaxy mass is limited by baryonic cooling within a dynamical time-scale and a minimum mass is given by the minimum virial temperature required for cooling to occur. This gives the Gaussian LF for giants and gas stripping then gives the Schechter function for dwarfs.   

\citet{driver:mgcmorph} find that the morphological classes E/S0, Sabc and Sd/Irr have distinct distributions relative to the bimodal population defined in the space of core $u-r$ colour and S\'ersic index $n$. The E/S0 and Sd/Irr correspond to the red and blue peaks respectively, whereas the Sabc spread across the two peaks. This is argued to be strong evidence that the bulges and discs are the fundamental two components to the bimodal population, not two separate galaxy types. Our ANN types contain few Sd/Irr due to the lack of these types in the training set, but our morphological LF supports this idea in the sense that the LFs for T types 1.5 through 3.5 (Sa through Sc corresponds to types 2--4) are essentially indistinguishable, in contrast to the E/S0 LFs which are distinctly brighter.

The bimodality in colour compared to that in morphology is also considered by B06 in the context of galaxy colour and ANN T-type morphology. There we find that whilst the distributions in colour are clearly bimodal and well-fit by a sum of two Gaussians, those in morphology are less so, with galaxies bridging the gap. This supports the non-fundamental nature of the Sabc types in this context.

Also discussed by, most recently, \citet{driver:mgcmorph} and \citet{ilbert:morphlfevoln} are blue spheroidal galaxies. These mimic faint ellipticals in morphology but are indistinguishable from blue galaxies by spectra. These galaxies could cause type contamination in the LFs, for example causing a steeper faint-end slope in the LF subdivided by $u-r$, as shown by \citeauthor{driver:mgcmorph}. Again this agrees with the trends seen in our LFs and therefore requires spectroscopic and morphological information to resolve.

Numerous studies \citep[e.g.][]{popesso:clusterlf, gonzalez:lfgroups} find that cluster LFs are better fit by a sum of a Gaussian at bright magnitudes and a Schechter at faint magnitudes, rather than a single Schechter function. The functions have a dip at $\mrlogh \sim -17$, caused by the flattening of the Gaussian before the power-law faint end kicks in. Field galaxies are consistent with a Schechter function. A hint of a similar dip is seen here, as might be expected as the sample encompasses all galaxy environments. The faint end slope is thought to be caused by numerous faint red galaxies which are present in clusters but not in the field. Our LFs do not probe deep enough to say anything definitive.

For the more Schechter-like results such as ours, the question is whether the LFs are as they appear, or whether the underlying types are Gaussian, appearing Schechter because type mixing makes the functions tend towards the overall LF, which is Schechter, or a superposition of Gaussian at the bright end and Schechter or power law at the faint end. The idea has previously been discussed in terms of the Hubble morphological types and the effects can be seen here, particularly in Fig. \ref{fig: ANN bivlf}. The LF for the early types could either be Schechter or Gaussian with a faint-end slope contaminated by dwarfs. In their sample which is essentially a subset of the one used here, \citet{nakamura:morphlf} use the concentration index to show that the sample is not contaminated with faint galaxies with soft cores, which will be dwarfs. Their concentration index LF is similar to their Hubble type LF. The same is seen here. 

The bimodality supports the idea that concentrated, early spectral type, red, high S\'ersic index galaxies correspond to the early morphological types. If this is the case then the Gaussian LF idea is also consistent with the LF plots bivariate with concentration index, eClass, reference frame colours and S\'ersic index. These all show LFs for the early types which could be Gaussian with a rise at faint magnitudes, rising to a Schechter function for later types. The idea is also supported by the LF bivariate with absolute Petrosian radius, which shows steeply declining faint-end slopes which become less steep for galaxies smaller than $\sim 2.5 \kpc$. 

Whatever processes cause the bimodality do not, however, show up in the LF subdivided by absolute effective surface brightness. However, it may well be the case that the sum of two bivariate distributions is still required and that, unlike the Cho{\l}oniewski function, the relation between the mean of the distribution on the surface brightness axis must be a non-linear function of absolute magnitude to fit the lack of variation in the peak surface brightness at bright magnitudes followed by dimming at fainter levels. At fainter levels the two distributions could diverge and thus reproduce the broadening seen. It may even be the case though that to get a formally good fit in the $\chi^2$ sense requires the distribution to be able to change width in the surface brightness axis as a function of absolute magnitude. If the Schechter function is also required to be a dual power law or Schechter plus Gaussian the number of parameters to fit rises to around 20. This may be impractical, at which point one would want to consider summing perhaps a greater number of simpler physically motivated functions, becoming more akin to semi-analytic models. These functions might be based on the Kormendy relation for bulges \citep{kormendy:relation}, Freeman's law for discs \citep{freeman:expprofile}, or the more recent relation of \citet{dejong:sbsdmlf}, and a relation for dwarf spheroids such as that in \citet{mateo:dwarfs}. Another possibility is that central surface brightness may better distinguish between spirals and ellipticals than the absolute effective surface brightness used here.

A useful way to connect observations to simulations is the conditional luminosity function, i.e. the luminosity distribution of galaxies in a dark matter halo. \citet{cooray:lstar} construct the Schechter form of the galaxy LF using a model in which the central massive galaxies in haloes are distributed with a lognormal scatter and the satellite galaxies, where the total galaxy luminosity is greater than the central galaxy, are a power law. This predicts that the faint end of the LF is determined by the galaxies in low mass haloes and should have a slope independent of colour selection or band. If this was combined with a prescription for predicting surface brightness then one might be able to predict the bivariate LFs presented here. This could be done by relating the absolute half-light radius to the dimensionless spin parameter of the dark matter halo \citep{peebles:spin}, as done by several studies \citep[e.g.][]{fall:galfmn, dalcanton:diskformation}. If the same dark matter haloes are used for both models then one could predict the bivariate brightness distribution, at least for discs. Relations of the haloes to other galaxy properties could similarly predict the other bivariate LFs, perhaps along the lines of e.g. \citet{mo:lfenvt}. However, our results show an LF faint end that does depend on the colour, so further comparison of observations and theory is clearly important.

The conclusion from various studies is that the two physical processes are likely to be merger to form spheroids and accretion on to discs to give the star formation. In the sense that we see the bimodal early-red versus late-blue population, our results support this conclusion. To test the ideas further requires bulge to disc decomposition and quantitative comparison with simulations incorporating the ideas discussed above, as well as the future work detailed below.

However, there is likely to be more to the story than simply mergers and accretion. For example, pseudobulges \citep[e.g.][]{kormendy:pseudobulges} resemble bulges in disc galaxies but are formed due to internal (secular) evolutionary processes. Thus the relative importance of secular evolution on the observables discussed here is a further avenue of investigation. Also, for example, none of the morphologies here take into account the presence of bars, which are important in a secular evolution sense, if not in larger scale environment \citep[e.g.][]{vandenbergh:bar}. There is then the question of more complex light profiles such as the core-S\'ersic model of \citet{graham:coresersic}, types that don't fit the classifications used here such as cD galaxies, higher redshifts, and so on.

Important followup work to the work presented in this paper is to more rigourously separate the dwarf and giant galaxies in the LF, to apply reliable bulge to disc decomposition and to fit functional forms to the two underlying distributions. More refined structural parameters than those used to generate the Main Galaxy Sample such as fully S\'ersic-based measurements using elliptical isophotes should be used. Here dust-obscuration is addressed by simply requiring that the axis ratio is less than that for an E7 elliptical galaxy, therefore excluding edge-on spiral discs that are highly reddened by internal extinction. Multiwavelength data could improve this, for example matches to SWIRE \citep{lonsdale:swire} data would be less affected. One could also match to GALEX \citep{martin:galex} data to extend to the UV for the star formation.

\section{Conclusions} \label{sec: conclusions}

The conclusions of this paper are as follows.\\

The luminosity function is measured bivariate with various galaxy properties from the SDSS and its associated value-added catalogues. The properties are eyeball morphological type, neural network morphological type, inverse concentration index $\ciinv$, the S\'ersic index of the light profile, absolute effective surface brightness, the reference frame colours $u-r$, $g-r$, $r-i$ and $r-z$, the absolute Petrosian 90\% radius of the galaxy, eClass PCA spectral type, the stellar mass and the $\Sigma_5$ surface density of galaxies. Some of the parameters plotted here are new in the context of a bivariate LF, and for those that have previously been studied, the trends seen are confirmed at high signal-to-noise. This is the largest number of parameters studied in one framework for a bivariate LF.

The SDSS is one of the best datasets currently available for this work, and the large sample combined with CCD photometry is ideal for precision bivariate LFs, enabling division of the galaxy population into numerous smaller samples, each of which is still highly statistically significant. In addition, the stepwise maximum likelihood estimator used is well-known and has been shown to be a good estimator. The magnitudes are K-corrected for each individual galaxy using the well-tested K-correct code written for the SDSS by \citet{blanton:kc}. 

The K-corrections do not take into account galaxy evolution or the effects of reddening due to dust, however we restrict the sample to galaxies at $z < 0.15$ and those with an isophotal axis ratio less than 10/3 to minimize these effects. Near-infrared datasets of comparable size to the SDSS should improve the situation.

The variation of the LF with all of the parameters is highly significant, both in shape and normalisation, except for $\Sigma_5$ where the LF does not change significantly in shape and the normalisation is arbitrary. 

The bivariate LFs are consistent with an early type, bright, concentrated, red population and a late type, faint, less concentrated, blue, star forming population. This idea of bimodality has been explored by others as it suggests two major underlying physical processes, which in agreement with other studies we conclude are likely to be mergers leading to bulges and accretion on to discs. The bimodality is considered further in B06 in the context of galaxy colour and morphology as a function of environment.

The various morphological measures are consistent, with the S\'ersic index showing the same broad trends as the $\ciinv$, eyeball and neural network morphological type.

The overall patterns of the LFs in the four colours $u-r$, $g-r$, $r-i$ and $r-z$ are similar, as are those for the other six from the SDSS $ugriz$ bands. The main change is the increased star formation in the bluer bands and the corresponding lack of bimodality in the redder bands.

The LF bivariate with absolute galactic radius shows a steeply declining faint-end slope which becomes flat then rising for the smallest galaxies, supporting the idea that it is the small galaxies which cause the LF to be shaped like a Schechter function.

The LF bivariate with galaxy stellar mass shows a tight correlation of higher mass for brighter magnitude, with Gaussian LFs, as expected as the luminosity is used in the calculation of the stellar mass.

The LF bivariate with $\Sigma_5$ surface density of galaxies shows little variation in shape over a large range in density from that in a cluster core to the field, being of power-law form throughout. Further work would be needed to quantify the variation that is seen.

No definitive conclusion is reached as to whether the underlying LFs for individual types are Gaussian or Schechter, due to the possibility of type mixing. This is particularly prominent in the LFs bivariate with eyeball morphological type assigned by the neural network, as those for the normal galaxies E, S0, and spirals might be expected to be Gaussian but if type mixing is present this causes them to tend towards the overall LF, which is well fit by a Schechter function. The idea is supported by the LFs bivariate with absolute galactic radius and to a lesser extent by S\'ersic index, concentration index, spectral type and colours (the bimodality suggests that these correspond to similar galaxy populations). However, the plots are also consistent with Schechter functions with more or less steeply declining faint-end slopes. Further work would be required to investigate the effect of blue spheroidals, which mimic faint red galaxies in morphology but have bluer colours.

It may be that the morphology of whole galaxies, particularly Sa--Sc spirals, is an intrinsically `fuzzy' measure, so that the Gaussian LFs will not be seen. However, the contamination may be from dwarf galaxies so an important step is to separate these from the normal galaxies and determine their LF separately. The LF by absolute galactic radius is a step towards this but the next step should be using the azimuthally averaged light profiles available for all galaxies in the SDSS. Similarly, further work would be required to see to what extent the faint-end upturns seen in some of the LF slices are real.

Some of the LFs are fitted with analytical functions, which are generally poor fits, probably due to this detail being beyond that reproducible by a simple function. An example is the fit of the Cho{\l}oniewski function to the bivariate brightness distribution in which the fit may need to also allow for a variation in the width of the Gaussian with surface brightness, and a possibly non-linear relation between mean absolute magnitude and mean surface brightness, which results in the addition of two parameters to the six already present. The poor fit is also seen in \citet{driver:mgcbbd}, where it is interpreted in terms of the diverging trends of bulges and discs with decreasing surface brightness. The fitting method used is easily extended to any of the other bivariate LFs presented here.

\section*{Acknowledgments}
We thank the referee for a useful report which greatly improved the paper.

Nick Ball thanks Osamu Nakamura for useful discussion on the JPG catalogue, Chris Miller and Michael Balogh for help with the VAC catalogue and Mike Blanton for help with the VAGC catalogue.

\noindent Nick Ball was funded by a PPARC studentship. NMB and RJB would like to acknowledge support from NASA through grants NAG5-12578 and NAG5-12580 as well as support through the NSF PACI Project.

Funding for the SDSS and SDSS-II has been provided by the Alfred P. Sloan Foundation, the Participating Institutions, the National Science Foundation, the U.S. Department of Energy, the National Aeronautics and Space Administration, the Japanese Monbukagakusho, the Max Planck Society, and the Higher Education Funding Council for England. The SDSS Web Site is http://www.sdss.org.

The SDSS is managed by the Astrophysical Research Consortium for the Participating Institutions. The Participating Institutions are the American Museum of Natural History, Astrophysical Institute Potsdam, University of Basel, Cambridge University, Case Western Reserve University, University of Chicago, Drexel University, Fermilab, the Institute for Advanced Study, the Japan Participation Group, Johns Hopkins University, the Joint Institute for Nuclear Astrophysics, the Kavli Institute for Particle Astrophysics and Cosmology, the Korean Scientist Group, the Chinese Academy of Sciences (LAMOST), Los Alamos National Laboratory, the Max-Planck-Institute for Astronomy (MPA), the Max-Planck-Institute for Astrophysics (MPIA), New Mexico State University, Ohio State University, University of Pittsburgh, University of Portsmouth, Princeton University, the United States Naval Observatory, and the University of Washington. 

This research has made use of NASA's Astrophysics Data System.


\begin{thebibliography}{}

\bibitem[\protect\citeauthoryear{{Abazajian} et~al.,}{{Abazajian}
  et~al.}{2003}]{abazajian:dr1}
{Abazajian} K.,  et~al., 2003, \aj, 126, 2081

\bibitem[\protect\citeauthoryear{{Abazajian} et~al.,}{{Abazajian}
  et~al.}{2004}]{abazajian:dr2}
{Abazajian} K.,  et~al., 2004, \aj, 128, 502

\bibitem[\protect\citeauthoryear{{Abell}}{{Abell}}{1962}]{abell:clusmem}
{Abell} G.~O.,  1962, in IAU Symp. 15: Problems of Extra-Galactic Research
  {Membership of Clusters of Galaxies}.
pp 213--+

\bibitem[\protect\citeauthoryear{{Adelman-McCarthy} et~al.,}{{Adelman-McCarthy}
   et~al.}{2006}]{adelmanmccarthy:dr4}
{Adelman-McCarthy} J.~K.,  et~al., 2006, \apjs, 162, 38

\bibitem[\protect\citeauthoryear{{Baldry}, {Glazebrook}, {Brinkmann}, {Ivezi{\'
  c}}, {Lupton}, {Nichol} \& {Szalay}}{{Baldry}
  et~al.}{2004}]{baldry:bimodalcmd}
{Baldry} I.~K.,  {Glazebrook} K.,  {Brinkmann} J.,  {Ivezi{\' c}} {\v Z}.,
  {Lupton} R.~H.,  {Nichol} R.~C.,    {Szalay} A.~S.,  2004, \apj, 600, 681

\bibitem[\protect\citeauthoryear{{Baldry} et~al.,}{{Baldry}
  et~al.}{2005}]{baldry:ulf}
{Baldry} I.~K., et~al., 2005, \mnras, 358, 441

\bibitem[\protect\citeauthoryear{{Ball}, {Loveday}, {Fukugita}, {Nakamura},
  {Okamura}, {Brinkmann} \& {Brunner}}{{Ball} et~al.}{2004}]{ball:ann}
{Ball} N.~M.,  {Loveday} J.,  {Fukugita} M.,  {Nakamura} O.,  {Okamura} S.,
  {Brinkmann} J.,    {Brunner} R.~J.,  2004, \mnras, 348, 1038

\bibitem[\protect\citeauthoryear{{Balogh}, {Baldry}, {Nichol}, {Miller},
  {Bower} \& {Glazebrook}}{{Balogh} et~al.}{2004}]{balogh:bimodallfenvt}
{Balogh} M.~L.,  {Baldry} I.~K.,  {Nichol} R.,  {Miller} C.,  {Bower} R.,
  {Glazebrook} K.,  2004, \apjl, 615, L101

\bibitem[\protect\citeauthoryear{{Benson}, {Bower}, {Frenk}, {Lacey}, {Baugh}
  \& {Cole}}{{Benson} et~al.}{2003}]{benson:lfshape}
{Benson} A.~J.,  {Bower} R.~G.,  {Frenk} C.~S.,  {Lacey} C.~G.,  {Baugh} C.~M.,
     {Cole} S.,  2003, \apj, 599, 38

\bibitem[\protect\citeauthoryear{{Benson}, {Cole}, {Frenk}, {Baugh} \&
  {Lacey}}{{Benson} et~al.}{2000}]{benson:bias}
{Benson} A.~J.,  {Cole} S.,  {Frenk} C.~S.,  {Baugh} C.~M.,    {Lacey} C.~G.,
  2000, \mnras, 311, 793

\bibitem[\protect\citeauthoryear{{Bernardi} et~al.,}{{Bernardi}
  et~al.}{2003}]{bernardi:elliptical2}
{Bernardi} M.,  et~al., 2003, \aj, 125, 1849

\bibitem[\protect\citeauthoryear{{Binggeli}, {Sandage} \& {Tammann}}{{Binggeli}
  et~al.}{1988}]{binggeli:lf}
{Binggeli} B.,  {Sandage} A.,    {Tammann} G.~A.,  1988, \araa, 26, 509

\bibitem[\protect\citeauthoryear{Bishop}{Bishop}{1995}]{bishop:ann}
Bishop C.~M.,  1995, Neural Networks for Pattern Recognition.
Oxford University Press, Oxford

\bibitem[\protect\citeauthoryear{{Blanton} et~al.,}{{Blanton}
  et~al.}{2001}]{blanton:edrlf}
{Blanton} M.~R.,  et~al., 2001, \aj, 121, 2358

\bibitem[\protect\citeauthoryear{{Blanton}, {Lin}, {Lupton}, {Maley}, {Young},
  {Zehavi} \& {Loveday}}{{Blanton} et~al.}{2003a}]{blanton:tiling}
{Blanton} M.~R.,  {Lin} H.,  {Lupton} R.~H.,  {Maley} F.~M.,  {Young} N.,
  {Zehavi} I.,    {Loveday} J.,  2003a, \aj, 125, 2276

\bibitem[\protect\citeauthoryear{{Blanton} et~al.,}{{Blanton}
  et~al.}{2003b}]{blanton:kc}
{Blanton} M.~R.,  et~al., 2003b, \aj, 125, 2348

\bibitem[\protect\citeauthoryear{{Blanton} et~al.,}{{Blanton}
  et~al.}{2003c}]{blanton:dr1lf}
{Blanton} M.~R.,  et~al., 2003c, \apj, 592, 819

\bibitem[\protect\citeauthoryear{{Blanton} et~al.,}{{Blanton}
  et~al.}{2005a}]{blanton:vagc}
{Blanton} M.~R.,  et~al., 2005a, \aj, 129, 2562

\bibitem[\protect\citeauthoryear{{Blanton}, {Eisenstein}, {Hogg}, {Schlegel} \&
  {Brinkmann}}{{Blanton} et~al.}{2005b}]{blanton:envtbbprops}
{Blanton} M.~R.,  {Eisenstein} D.,  {Hogg} D.~W.,  {Schlegel} D.~J.,
  {Brinkmann} J.,  2005b, \apj, 629, 143

\bibitem[\protect\citeauthoryear{{Blanton}, {Lupton}, {Schlegel}, {Strauss},
  {Brinkmann}, {Fukugita} \& {Loveday}}{{Blanton}
  et~al.}{2005c}]{blanton:lflowl}
{Blanton} M.~R.,  {Lupton} R.~H.,  {Schlegel} D.~J.,  {Strauss} M.~A.,
  {Brinkmann} J.,  {Fukugita} M.,    {Loveday} J.,  2005c, \apj, 631, 208

\bibitem[\protect\citeauthoryear{{Boyce} \& {Phillipps}}{{Boyce} \&
  {Phillipps}}{1995}]{boyce:bbd}
{Boyce} P.~J.,  {Phillipps} S.,  1995, \aap, 296, 26

\bibitem[\protect\citeauthoryear{{Budav{\'a}ri} et~al.,}{{Budav{\'a}ri}
  et~al.}{2005}]{budavari:galexlf}
{Budav{\'a}ri} T.,  et~al., 2005, \apjl, 619, L31

\bibitem[\protect\citeauthoryear{{Cho{\l}oniewski}}{{Cho{\l}oniewski}}{1985}]{%
choloniewski:bivlf}
{Cho{\l}oniewski} J.,  1985, \mnras, 214, 197

\bibitem[\protect\citeauthoryear{{Cole} et~al.,}{{Cole}
  et~al.}{2001}]{cole:nirlf}
{Cole} S.,  et~al., 2001, \mnras, 326, 255

\bibitem[\protect\citeauthoryear{{Colless} et~al.,}{{Colless}
  et~al.}{2001}]{colless:2df}
{Colless} M.,  et~al., 2001, \mnras, 328, 1039

\bibitem[\protect\citeauthoryear{{Colless} et~al.,}{{Colless}
  et~al.}{2003}]{colless:2dffinal}
{Colless} M.,  et~al., 2003, Preprint: astro-ph/0306581

\bibitem[\protect\citeauthoryear{{Connolly} et~al.,}{{Connolly}
  et~al.}{2002}]{connolly:edrcf}
{Connolly} A.~J.,  et~al., 2002, \apj, 579, 42

\bibitem[\protect\citeauthoryear{{Connolly} \& {Szalay}}{{Connolly} \&
  {Szalay}}{1999}]{connolly:eclass}
{Connolly} A.~J.,  {Szalay} A.~S.,  1999, \aj, 117, 2052

\bibitem[\protect\citeauthoryear{{Connolly}, {Szalay}, {Bershady}, {Kinney} \&
  {Calzetti}}{{Connolly} et~al.}{1995}]{connolly:orthogonal}
{Connolly} A.~J.,  {Szalay} A.~S.,  {Bershady} M.~A.,  {Kinney} A.~L.,
  {Calzetti} D.,  1995, \aj, 110, 1071

\bibitem[\protect\citeauthoryear{{Cooray} \& {Milosavljevi{\'c}}}{{Cooray} \&
  {Milosavljevi{\'c}}}{2005}]{cooray:lstar}
{Cooray} A.,  {Milosavljevi{\'c}} M.,  2005, \apjl, 627, L89

\bibitem[\protect\citeauthoryear{{Cross} \& {Driver}}{{Cross} \&
  {Driver}}{2002}]{cross:bbd}
{Cross} N.,  {Driver} S.~P.,  2002, \mnras, 329, 579

\bibitem[\protect\citeauthoryear{{Cross} et~al.,}{{Cross}
  et~al.}{2001}]{cross:ldens}
{Cross} N.,  et~al., 2001, \mnras, 324, 825

\bibitem[\protect\citeauthoryear{{Croton} et~al.,}{{Croton}
  et~al.}{2005}]{croton:lfenvt}
{Croton} D.~J.,  et~al., 2005, \mnras, 356, 1155

\bibitem[\protect\citeauthoryear{{Dalcanton}, {Spergel} \&
  {Summers}}{{Dalcanton} et~al.}{1997}]{dalcanton:diskformation}
{Dalcanton} J.~J.,  {Spergel} D.~N.,    {Summers} F.~J.,  1997, \apj, 482, 659

\bibitem[\protect\citeauthoryear{{Davis} \& {Huchra}}{{Davis} \&
  {Huchra}}{1982}]{davis:cfadens}
{Davis} M.,  {Huchra} J.,  1982, \apj, 254, 437

\bibitem[\protect\citeauthoryear{{de Jong} \& {Lacey}}{{de Jong} \&
  {Lacey}}{2000}]{dejong:sbsdmlf}
{de Jong} R.~S.,  {Lacey} C.,  2000, \apj, 545, 781

\bibitem[\protect\citeauthoryear{{de Jong}, {Simard}, {Davies}, {Saglia},
  {Burstein}, {Colless}, {McMahan} \& {Wegner}}{{de Jong}
  et~al.}{2004}]{dejong:bdearly}
{de Jong} R.~S.,  {Simard} L.,  {Davies} R.~L.,  {Saglia} R.~P.,  {Burstein}
  D.,  {Colless} M.,  {McMahan} R.,    {Wegner} G.,  2004, \mnras, 355, 1155

\bibitem[\protect\citeauthoryear{{de Lapparent}}{{de
  Lapparent}}{2003}]{delapparent:morphlf}
{de Lapparent} V.,  2003, \aap, 408, 845

\bibitem[\protect\citeauthoryear{{de Lapparent}, {Arnouts}, {Galaz} \&
  {Bardelli}}{{de Lapparent} et~al.}{2004}]{delapparent:spirallfevoln}
{de Lapparent} V.,  {Arnouts} S.,  {Galaz} G.,    {Bardelli} S.,  2004, \aap,
  422, 841

\bibitem[\protect\citeauthoryear{{de Lapparent}, {Galaz}, {Bardelli} \&
  {Arnouts}}{{de Lapparent} et~al.}{2003}]{delapparent:speclf}
{de Lapparent} V.,  {Galaz} G.,  {Bardelli} S.,    {Arnouts} S.,  2003, \aap,
  404, 831

\bibitem[\protect\citeauthoryear{{de Vaucouleurs}}{{de
  Vaucouleurs}}{1948}]{devaucouleurs:devprofile}
{de Vaucouleurs} G.,  1948, Annales d'Astrophysique, 11, 247

\bibitem[\protect\citeauthoryear{{de Vaucouleurs}}{{de
  Vaucouleurs}}{1959}]{devaucouleurs:ttype}
{de Vaucouleurs} G.,  1959, Handbuch der Physik, 53, 275

\bibitem[\protect\citeauthoryear{{Dressler}}{{Dressler}}{1980}]{dressler:morph%
density}
{Dressler} A.,  1980, \apj, 236, 351

\bibitem[\protect\citeauthoryear{{Drinkwater}, {Jones}, {Gregg} \&
  {Phillipps}}{{Drinkwater} et~al.}{2000}]{drinkwater:ucd}
{Drinkwater} M.~J.,  {Jones} J.~B.,  {Gregg} M.~D.,    {Phillipps} S.,  2000,
  Publications of the Astronomical Society of Australia, 17, 227

\bibitem[\protect\citeauthoryear{{Driver}}{{Driver}}{2004}]{driver:beyondlf}
{Driver} S.,  2004, Publications of the Astronomical Society of Australia, 21,
  344

\bibitem[\protect\citeauthoryear{{Driver} et~al.,}{{Driver}
  et~al.}{2006}]{driver:mgcmorph}
{Driver} S.~P.,  et~al., 2006, \mnras, 368, 414

\bibitem[\protect\citeauthoryear{{Driver}, {Liske}, {Cross}, {De Propris} \&
  {Allen}}{{Driver} et~al.}{2005}]{driver:mgcbbd}
{Driver} S.~P.,  {Liske} J.,  {Cross} N.~J.~G.,  {De Propris} R.,    {Allen}
  P.~D.,  2005, \mnras, 360, 81

\bibitem[\protect\citeauthoryear{{Efstathiou}, {Ellis} \&
  {Peterson}}{{Efstathiou} et~al.}{1988}]{efstathiou:swml}
{Efstathiou} G.,  {Ellis} R.~S.,    {Peterson} B.~A.,  1988, \mnras, 232, 431

\bibitem[\protect\citeauthoryear{{Eisenstein} et~al.,}{{Eisenstein}
  et~al.}{2001}]{eisenstein:lrgsample}
{Eisenstein} D.~J.,  et~al., 2001, \aj, 122, 2267

\bibitem[\protect\citeauthoryear{{Eke}, {Baugh}, {Cole}, {Frenk}, {King} \&
  {Peacock}}{{Eke} et~al.}{2004}]{eke:kband}
{Eke} V.~R.,  {Baugh} C.~M.,  {Cole} S.,  {Frenk} C.~S.,  {King} H.~M.,
  {Peacock} J.~A.,  2004, Preprint: astro-ph/0412049

\bibitem[\protect\citeauthoryear{{Fall} \& {Efstathiou}}{{Fall} \&
  {Efstathiou}}{1980}]{fall:galfmn}
{Fall} S.~M.,  {Efstathiou} G.,  1980, \mnras, 193, 189

\bibitem[\protect\citeauthoryear{{Freeman}}{{Freeman}}{1970}]{freeman:expprofi%
le}
{Freeman} K.~C.,  1970, \apj, 160, 811

\bibitem[\protect\citeauthoryear{{Fukugita}, {Ichikawa}, {Gunn}, {Doi},
  {Shimasaku} \& {Schneider}}{{Fukugita}
  et~al.}{1996}]{fukugita:sdssphotometry}
{Fukugita} M.,  {Ichikawa} T.,  {Gunn} J.~E.,  {Doi} M.,  {Shimasaku} K.,
  {Schneider} D.~P.,  1996, \aj, 111, 1748

\bibitem[\protect\citeauthoryear{{Glazebrook} et~al.,}{{Glazebrook}
  et~al.}{2003}]{glazebrook:cosmicspectrum}
{Glazebrook} K.,  et~al., 2003, \apj, 587, 55

\bibitem[\protect\citeauthoryear{{Gonz{\'a}lez}, {Lares}, {Lambas} \&
  {Valotto}}{{Gonz{\'a}lez} et~al.}{2006}]{gonzalez:lfgroups}
{Gonz{\'a}lez} R.~E.,  {Lares} M.,  {Lambas} D.~G.,    {Valotto} C.,  2006,
  \aap, 445, 51

\bibitem[\protect\citeauthoryear{{Graham} \& {Driver}}{{Graham} \&
  {Driver}}{2005}]{graham:sersic}
{Graham} A.~W.,  {Driver} S.~P.,  2005, \pasa, 22, 118

\bibitem[\protect\citeauthoryear{{Graham}, {Driver}, {Petrosian}, {Conselice},
  {Bershady}, {Crawford} \& {Goto}}{{Graham} et~al.}{2005}]{graham:totalmag}
{Graham} A.~W.,  {Driver} S.~P.,  {Petrosian} V.,  {Conselice} C.~J.,
  {Bershady} M.~A.,  {Crawford} S.~M.,    {Goto} T.,  2005, \aj, 130, 1535

\bibitem[\protect\citeauthoryear{{Graham}, {Erwin}, {Trujillo} \& {Asensio
  Ramos}}{{Graham} et~al.}{2003}]{graham:coresersic}
{Graham} A.~W.,  {Erwin} P.,  {Trujillo} I.,    {Asensio Ramos} A.,  2003, \aj,
  125, 2951

\bibitem[\protect\citeauthoryear{{Gunn} et~al.,}{{Gunn}
  et~al.}{1998}]{gunn:sdsscamera}
{Gunn} J.~E.,  et~al., 1998, \aj, 116, 3040

\bibitem[\protect\citeauthoryear{{Gunn} et~al.,}{{Gunn}
  et~al.}{2006}]{gunn:sdsstelescope}
{Gunn} J.~E.,  et~al., 2006, \aj, 131, 2332

\bibitem[\protect\citeauthoryear{{Hamilton}}{{Hamilton}}{1993}]{hamilton:cf}
{Hamilton} A.~J.~S.,  1993, \apj, 417, 19

\bibitem[\protect\citeauthoryear{{Hayward}, {Irwin} \& {Bregman}}{{Hayward}
  et~al.}{2005}]{hayward:unimportance}
{Hayward} C.~C.,  {Irwin} J.~A.,    {Bregman} J.~N.,  2005, \apj, 635, 827

\bibitem[\protect\citeauthoryear{{Hogg}, {Finkbeiner}, {Schlegel} \&
  {Gunn}}{{Hogg} et~al.}{2001}]{hogg:sdssmt}
{Hogg} D.~W.,  {Finkbeiner} D.~P.,  {Schlegel} D.~J.,    {Gunn} J.~E.,  2001,
  \aj, 122, 2129

\bibitem[\protect\citeauthoryear{{Holmberg}}{{Holmberg}}{1974}]{holmberg:group%
s}
{Holmberg} E.,  1974, Arkiv for Astronomi, 5, 305

\bibitem[\protect\citeauthoryear{{Hoyle}, {Rojas}, {Vogeley} \&
  {Brinkmann}}{{Hoyle} et~al.}{2005}]{hoyle:voidlf}
{Hoyle} F.,  {Rojas} R.~R.,  {Vogeley} M.~S.,    {Brinkmann} J.,  2005, \apj,
  620, 618

\bibitem[\protect\citeauthoryear{{Hubble}}{{Hubble}}{1936a}]{hubble:realm}
{Hubble} E.~P.,  1936a, Yale University Press

\bibitem[\protect\citeauthoryear{{Hubble}}{{Hubble}}{1936b}]{hubble:lfnebulaes%
tars}
{Hubble} E.~P.,  1936b, \apj, 84, 158

\bibitem[\protect\citeauthoryear{{Hubble}}{{Hubble}}{1936c}]{hubble:lfnebulaev%
elmag}
{Hubble} E.~P.,  1936c, \apj, 84, 270

\bibitem[\protect\citeauthoryear{{Ilbert} et~al.,}{{Ilbert}
  et~al.}{2006}]{ilbert:morphlfevoln}
{Ilbert} O.,  et~al., 2006, \aap, 453, 809

\bibitem[\protect\citeauthoryear{{Ivezi{\' c}} et~al.,}{{Ivezi{\' c}}
  et~al.}{2004}]{ivezic:sdssdata}
{Ivezi{\' c}} {\v Z}.,  et~al., 2004, Astronomische Nachrichten, 325, 583

\bibitem[\protect\citeauthoryear{{Jerjen} \& {Tammann}}{{Jerjen} \&
  {Tammann}}{1997}]{jerjen:morphlf}
{Jerjen} H.,  {Tammann} G.~A.,  1997, \aap, 321, 713

\bibitem[\protect\citeauthoryear{{Jones}, {Peterson}, {Colless} \&
  {Saunders}}{{Jones} et~al.}{2006}]{jones:6dflf}
{Jones} D.~H.,  {Peterson} B.~A.,  {Colless} M.,    {Saunders} W.,  2006,
  \mnras, 369, 25

\bibitem[\protect\citeauthoryear{{Kauffmann} et~al.,}{{Kauffmann}
  et~al.}{2003}]{kauffmann:masssfh}
{Kauffmann} G.,  et~al., 2003, \mnras, 341, 33

\bibitem[\protect\citeauthoryear{Kendall \& Stuart}{Kendall \&
  Stuart}{1961}]{kendall:stat}
Kendall M.~G.,  Stuart A.,  1961, The Advanced Theory of Statistics.
Vol.~2, Griffin \& Griffin, London

\bibitem[\protect\citeauthoryear{Kochanek, Pahre \& Falco}{Kochanek
  et~al.}{2000}]{kochanek:inconsistencies}
Kochanek C.~S.,  Pahre M.~A.,    Falco E.~E.,  2000, Preprint, astro-ph/0011458

\bibitem[\protect\citeauthoryear{{Kochanek}, {Pahre}, {Falco}, {Huchra},
  {Mader}, {Jarrett}, {Chester}, {Cutri} \& {Schneider}}{{Kochanek}
  et~al.}{2001}]{kochanek:kbandlf}
{Kochanek} C.~S.,  {Pahre} M.~A.,  {Falco} E.~E.,  {Huchra} J.~P.,  {Mader} J.,
   {Jarrett} T.~H.,  {Chester} T.,  {Cutri} R.,    {Schneider} S.~E.,  2001,
  \apj, 560, 566

\bibitem[\protect\citeauthoryear{{Kormendy}}{{Kormendy}}{1977}]{kormendy:relat%
ion}
{Kormendy} J.,  1977, \apj, 218, 333

\bibitem[\protect\citeauthoryear{{Kormendy} \& {Kennicutt}}{{Kormendy} \&
  {Kennicutt}}{2004}]{kormendy:pseudobulges}
{Kormendy} J.,  {Kennicutt} R.~C.,  2004, \araa, 42, 603

\bibitem[\protect\citeauthoryear{{Kuehn} \& {Ryden}}{{Kuehn} \&
  {Ryden}}{2005}]{kuehn:shape}
{Kuehn} F.,  {Ryden} B.~S.,  2005, \apj, 634, 1032

\bibitem[\protect\citeauthoryear{Lagarias, Reeds, Wright \& Wright}{Lagarias
  et~al.}{1998}]{lagarias:simplex}
Lagarias J.,  Reeds J.~A.,  Wright M.~H.,    Wright P.~E.,  1998, SIAM Journal
  of Optimization, 9, 112

\bibitem[\protect\citeauthoryear{{Lahav}, {Naim}, {Sodr{\' e}} \&
  {Storrie-Lombardi}}{{Lahav} et~al.}{1996}]{lahav:annmethods}
{Lahav} O.,  {Naim} A.,  {Sodr{\' e}} L.,    {Storrie-Lombardi} M.~C.,  1996,
  \mnras, 283, 207

\bibitem[\protect\citeauthoryear{{Limber}}{{Limber}}{1953}]{limber:deprojectio%
n}
{Limber} D.~N.,  1953, \apj, 117, 134

\bibitem[\protect\citeauthoryear{{Liske}, {Lemon}, {Driver}, {Cross} \&
  {Couch}}{{Liske} et~al.}{2003}]{liske:mgclf}
{Liske} J.,  {Lemon} D.~J.,  {Driver} S.~P.,  {Cross} N.~J.~G.,    {Couch}
  W.~J.,  2003, \mnras, 344, 307

\bibitem[\protect\citeauthoryear{{Lonsdale} et~al.,}{{Lonsdale}
  et~al.}{2003}]{lonsdale:swire}
{Lonsdale} C.~J.,  et~al., 2003, \pasp, 115, 897

\bibitem[\protect\citeauthoryear{{Loveday}}{{Loveday}}{2000}]{loveday:kbandlf}
{Loveday} J.,  2000, \mnras, 312, 557

\bibitem[\protect\citeauthoryear{{Loveday}}{{Loveday}}{2004}]{loveday:lfevol}
{Loveday} J.,  2004, \mnras, 347, 601

\bibitem[\protect\citeauthoryear{{Lupton}}{{Lupton}}{2006}]{lupton:deblender}
{Lupton} R.~H.,  2006, AJ, submitted

\bibitem[\protect\citeauthoryear{{Lupton}, {Gunn}, {Ivezi{\' c}}, {Knapp},
  {Kent} \& {Yasuda}}{{Lupton} et~al.}{2001}]{lupton:sdssphoto}
{Lupton} R.~H.,  {Gunn} J.~E.,  {Ivezi{\' c}} Z.,  {Knapp} G.~R.,  {Kent} S.,
   {Yasuda} N.,  2001, in ASP Conf. Ser. 238: Astronomical Data Analysis
  Software and Systems X {The SDSS Imaging Pipelines}

\bibitem[\protect\citeauthoryear{{Lupton}, {Gunn} \& {Szalay}}{{Lupton}
  et~al.}{1999}]{lupton:asinhmag}
{Lupton} R.~H.,  {Gunn} J.~E.,    {Szalay} A.~S.,  1999, \aj, 118, 1406

\bibitem[\protect\citeauthoryear{{Madgwick} et~al.,}{{Madgwick}
  et~al.}{2002}]{madgwick:speclf}
{Madgwick} D.~S.,  et~al., 2002, \mnras, 333, 133

\bibitem[\protect\citeauthoryear{{Martin} et~al.,}{{Martin}
  et~al.}{2005}]{martin:galex}
{Martin} D.~C.,  et~al., 2005, \apjl, 619, L1

\bibitem[\protect\citeauthoryear{{Mateo}}{{Mateo}}{1998}]{mateo:dwarfs}
{Mateo} M.~L.,  1998, \araa, 36, 435

\bibitem[\protect\citeauthoryear{{Mathewson} \& {Ford}}{{Mathewson} \&
  {Ford}}{1996}]{mathewson:mfb2}
{Mathewson} D.~S.,  {Ford} V.~L.,  1996, \apjs, 107, 97

\bibitem[\protect\citeauthoryear{{Mathewson}, {Ford} \& {Buchhorn}}{{Mathewson}
  et~al.}{1992}]{mathewson:mfb}
{Mathewson} D.~S.,  {Ford} V.~L.,    {Buchhorn} M.,  1992, \apjs, 81, 413

\bibitem[\protect\citeauthoryear{{Mercurio} et~al.,}{{Mercurio}
  et~al.}{2006}]{mercurio:superclusterlf}
{Mercurio} A.,  et~al., 2006, \mnras, pp 314--+

\bibitem[\protect\citeauthoryear{{Miyaji}, {Hasinger} \& {Schmidt}}{{Miyaji}
  et~al.}{2000}]{miyaji:xraylf}
{Miyaji} T.,  {Hasinger} G.,    {Schmidt} M.,  2000, \aap, 353, 25

\bibitem[\protect\citeauthoryear{{Mo}, {Yang}, {van den Bosch} \& {Jing}}{{Mo}
  et~al.}{2004}]{mo:lfenvt}
{Mo} H.~J.,  {Yang} X.,  {van den Bosch} F.~C.,    {Jing} Y.~P.,  2004, \mnras,
  349, 205

\bibitem[\protect\citeauthoryear{{Nakamura}, {Fukugita}, {Yasuda}, {Loveday},
  {Brinkmann}, {Schneider}, {Shimasaku} \& {SubbaRao}}{{Nakamura}
  et~al.}{2003}]{nakamura:morphlf}
{Nakamura} O.,  {Fukugita} M.,  {Yasuda} N.,  {Loveday} J.,  {Brinkmann} J.,
  {Schneider} D.~P.,  {Shimasaku} K.,    {SubbaRao} M.,  2003, \aj, 125, 1682

\bibitem[\protect\citeauthoryear{Nelder \& Mead}{Nelder \&
  Mead}{1965}]{nelder:simplex}
Nelder J.~A.,  Mead R.,  1965, Computer Journal, 7, 308

\bibitem[\protect\citeauthoryear{{Norberg} et~al.,}{{Norberg}
  et~al.}{2002}]{norberg:lf}
{Norberg} P.,  et~al., 2002, \mnras, 336, 907

\bibitem[\protect\citeauthoryear{{Peebles}}{{Peebles}}{1969}]{peebles:spin}
{Peebles} P.~J.~E.,  1969, \apj, 155, 393

\bibitem[\protect\citeauthoryear{{Petrosian}}{{Petrosian}}{1976}]{petrosian:pe%
tromag}
{Petrosian} V.,  1976, \apjl, 209, L1

\bibitem[\protect\citeauthoryear{{Phillipps} \& {Disney}}{{Phillipps} \&
  {Disney}}{1986}]{phillipps:selection}
{Phillipps} S.,  {Disney} M.,  1986, \mnras, 221, 1039

\bibitem[\protect\citeauthoryear{{Phillipps}, {Drinkwater}, {Gregg} \&
  {Jones}}{{Phillipps} et~al.}{2001}]{phillipps:ucd}
{Phillipps} S.,  {Drinkwater} M.~J.,  {Gregg} M.~D.,    {Jones} J.~B.,  2001,
  \apj, 560, 201

\bibitem[\protect\citeauthoryear{{Pier}, {Munn}, {Hindsley}, {Hennessy},
  {Kent}, {Lupton} \& {Ivezi{\' c}}}{{Pier} et~al.}{2003}]{pier:sdssastrometry}
{Pier} J.~R.,  {Munn} J.~A.,  {Hindsley} R.~B.,  {Hennessy} G.~S.,  {Kent}
  S.~M.,  {Lupton} R.~H.,    {Ivezi{\' c}} {\v Z}.,  2003, \aj, 125, 1559

\bibitem[\protect\citeauthoryear{{Popesso}, {B{\"o}hringer}, {Romaniello} \&
  {Voges}}{{Popesso} et~al.}{2005}]{popesso:clusterlf}
{Popesso} P.,  {B{\"o}hringer} H.,  {Romaniello} M.,    {Voges} W.,  2005,
  \aap, 433, 415

\bibitem[\protect\citeauthoryear{{Press} \& {Schechter}}{{Press} \&
  {Schechter}}{1974}]{press:structfmn}
{Press} W.~H.,  {Schechter} P.,  1974, \apj, 187, 425

\bibitem[\protect\citeauthoryear{{Ranalli}, {Comastri} \& {Setti}}{{Ranalli}
  et~al.}{2005}]{ranalli:xraylf}
{Ranalli} P.,  {Comastri} A.,    {Setti} G.,  2005, \aap, 440, 23

\bibitem[\protect\citeauthoryear{{Read} \& {Trentham}}{{Read} \&
  {Trentham}}{2005}]{read:massfn}
{Read} J.~I.,  {Trentham} N.,  2005, Royal Society of London Philosophical
  Transactions Series A, 363, 2693

\bibitem[\protect\citeauthoryear{{Reda}, {Forbes}, {Beasley}, {O'Sullivan} \&
  {Goudfrooij}}{{Reda} et~al.}{2004}]{reda:earlylf}
{Reda} F.~M.,  {Forbes} D.~A.,  {Beasley} M.~A.,  {O'Sullivan} E.~J.,
  {Goudfrooij} P.,  2004, \mnras, 354, 851

\bibitem[\protect\citeauthoryear{{Richards} et~al.,}{{Richards}
  et~al.}{2002}]{richards:qsosample}
{Richards} G.~T.,  et~al., 2002, \aj, 123, 2945

\bibitem[\protect\citeauthoryear{{Sadler} et~al.,}{{Sadler}
  et~al.}{2002}]{sadler:radiolf}
{Sadler} E.~M.,  et~al., 2002, \mnras, 329, 227

\bibitem[\protect\citeauthoryear{{Sandage}, {Binggeli} \& {Tammann}}{{Sandage}
  et~al.}{1985}]{sandage:virgolf}
{Sandage} A.,  {Binggeli} B.,    {Tammann} G.~A.,  1985, \aj, 90, 1759

\bibitem[\protect\citeauthoryear{{Schaeffer} \& {Silk}}{{Schaeffer} \&
  {Silk}}{1988}]{schaeffer:bimodallf}
{Schaeffer} R.,  {Silk} J.,  1988, \aap, 203, 273

\bibitem[\protect\citeauthoryear{{Schechter}}{{Schechter}}{1976}]{schechter:lf}
{Schechter} P.,  1976, \apj, 203, 297

\bibitem[\protect\citeauthoryear{{Schlegel}, {Finkbeiner} \&
  {Davis}}{{Schlegel} et~al.}{1998}]{schlegel:dustmaps}
{Schlegel} D.~J.,  {Finkbeiner} D.~P.,    {Davis} M.,  1998, \apj, 500, 525

\bibitem[\protect\citeauthoryear{{S\'ersic}}{{S\'ersic}}{1968}]{sersic:austral%
es}
{S\'ersic} J.~L.,  1968, {Atlas de galaxias australes}.
Cordoba, Argentina: Observatorio Astronomico, 1968

\bibitem[\protect\citeauthoryear{{Shen}, {Mo}, {White}, {Blanton}, {Kauffmann},
  {Voges}, {Brinkmann} \& {Csabai}}{{Shen} et~al.}{2003}]{shen:sdssgalsizes}
{Shen} S.,  {Mo} H.~J.,  {White} S.~D.~M.,  {Blanton} M.~R.,  {Kauffmann} G.,
  {Voges} W.,  {Brinkmann} J.,    {Csabai} I.,  2003, \mnras, 343, 978

\bibitem[\protect\citeauthoryear{{Shimasaku} et~al.,}{{Shimasaku}
  et~al.}{2001}]{shimasaku:brightgal}
{Shimasaku} K.,  et~al., 2001, \aj, 122, 1238

\bibitem[\protect\citeauthoryear{{Smith} et~al.,}{{Smith}
  et~al.}{2002}]{smith:ugrizstars}
{Smith} J.~A.,  et~al., 2002, \aj, 123, 2121

\bibitem[\protect\citeauthoryear{{Sodr\'e} \& {Lahav}}{{Sodr\'e} \&
  {Lahav}}{1993}]{sodre:esolvbivlf}
{Sodr\'e} L.~J.,  {Lahav} O.,  1993, \mnras, 260, 285

\bibitem[\protect\citeauthoryear{{Stocke}, {Keeney}, {Lewis}, {Epps} \&
  {Schild}}{{Stocke} et~al.}{2004}]{stocke:earlytypelf}
{Stocke} J.~T.,  {Keeney} B.~A.,  {Lewis} A.~D.,  {Epps} H.~W.,    {Schild}
  R.~E.,  2004, \aj, 127, 1336

\bibitem[\protect\citeauthoryear{{Stoughton} et~al.,}{{Stoughton}
  et~al.}{2002}]{stoughton:edr}
{Stoughton} C.,  et~al., 2002, \aj, 123, 485

\bibitem[\protect\citeauthoryear{{Strauss} et~al.,}{{Strauss}
  et~al.}{2002}]{strauss:mainsample}
{Strauss} M.~A.,  et~al., 2002, \aj, 124, 1810

\bibitem[\protect\citeauthoryear{{Takeuchi}, {Yoshikawa} \& {Ishii}}{{Takeuchi}
  et~al.}{2000}]{takeuchi:lfests}
{Takeuchi} T.~T.,  {Yoshikawa} K.,    {Ishii} T.~T.,  2000, \apjs, 129, 1

\bibitem[\protect\citeauthoryear{{Treyer} et~al.,}{{Treyer}
  et~al.}{2005}]{treyer:galexlf}
{Treyer} M.,  et~al., 2005, \apjl, 619, L19

\bibitem[\protect\citeauthoryear{{Tucker} et~al.,}{{Tucker}
  et~al.}{2006}]{tucker:calibration}
{Tucker} D.,  et~al., 2006, Astronomische Nachrichten, in press

\bibitem[\protect\citeauthoryear{{van den Bergh}}{{van den
  Bergh}}{2002}]{vandenbergh:bar}
{van den Bergh} S.,  2002, \aj, 124, 782

\bibitem[\protect\citeauthoryear{{Vincent} \& {Ryden}}{{Vincent} \&
  {Ryden}}{2005}]{vincent:shape}
{Vincent} R.~A.,  {Ryden} B.~S.,  2005, \apj, 623, 137

\bibitem[\protect\citeauthoryear{{Wyder} et~al.,}{{Wyder}
  et~al.}{2005}]{wyder:galexuvlf}
{Wyder} T.~K.,  et~al., 2005, \apjl, 619, L15

\bibitem[\protect\citeauthoryear{{Yip} et~al.,}{{Yip}
  et~al.}{2004}]{yip:eclass}
{Yip} C.~W.,  et~al., 2004, \aj, 128, 585

\bibitem[\protect\citeauthoryear{{York} et~al.,}{{York}
  et~al.}{2000}]{york:sdss}
{York} D.~G.,  et~al., 2000, \aj, 120, 1579

\bibitem[\protect\citeauthoryear{{Zehavi} et~al.,}{{Zehavi}
  et~al.}{2002}]{zehavi:edrclustering}
{Zehavi} I.,  et~al., 2002, \apj, 571, 172

\bibitem[\protect\citeauthoryear{{Zehavi} et~al.,}{{Zehavi}
  et~al.}{2004a}]{zehavi:gcfpowerlaw}
{Zehavi} I.,  et~al., 2004a, \apj, 608, 16

\bibitem[\protect\citeauthoryear{{Zehavi} et~al.,}{{Zehavi}
  et~al.}{2004b}]{zehavi:lcgcf}
{Zehavi} I.,  et~al., 2004b, \apj, 630, 1

\bibitem[\protect\citeauthoryear{{Zibetti}, {White} \& {Brinkmann}}{{Zibetti}
  et~al.}{2004}]{zibetti:edgeon}
{Zibetti} S.,  {White} S.~D.~M.,    {Brinkmann} J.,  2004, \mnras, 347, 556

\end{thebibliography}


\appendix

\section{Completeness Maps} \label{app: completeness maps}

For each result in which the second LF parameter does not require spectroscopy, i.e. $\tjpg$, $\tann$, $\ciinv$, and $n$, the completeness maps for the bins used in the LF SWML estimate are shown. The parameters are therefore selected with the explicit cuts implied by the extent of their axes on these maps. For $\sbr$ we show the completeness as a function of apparent SB. Maps for the other parameters are shown as a function of apparent magnitude. With the exception of $\tjpg$, which is selected from a smaller area with much higher spectroscopic coverage, the overall completeness will generally not exceed the ratio of the spectroscopic areal coverage to that of the imaging over the whole survey. This is about 70\%, from 4,783 or 4,681 $\sqdeg$ versus 6,851 or 6,670 $\sqdeg$ for the VAGC and the CAS respectively. In general, the completeness will be lower, due to further objects being masked out for the various reasons described above. The maps give an indication of the possible extent of biases in the LFs due to incompleteness in spectroscopy.

Whilst these maps show the completeness in spectroscopy compared to selected targets, they do not show incompletenesses in the imaging compared to the true galaxy distribution. These will arise if either the object is not detected, or it is masked out from the photometry that is compared to the spectroscopy in the maps.

The photometric incompleteness due to surface brightness selection effects is investigated by \citet{blanton:lflowl} for the ranges in apparent surface brightness and magnitude $18 < \sbrapp < 24.5$ and $14 < r < 17.5$. They process simulated galaxies with the same software used to process the real DR2 imaging data. The effects of seeing, noise and sky subtraction are included. At low surface brightness, they find that the overall completeness drops below 90\% around $\sbrapp > 22$, reaching 50\% at $\sbrapp = 23.4$ and 0 at $\sbrapp$ = 24.3. This drop is fairly uniform over the whole range in $r$. At the bright end, there is slight incompleteness (90\%) for dim high surface brightness objects ($\sbrapp < 19$ and $r > 17$), ascribed to the star/galaxy separator misclassifying the objects as stars in marginal seeing. There is more substantial incompleteness for bright high surface brightness objects, the overall completeness dropping to 80\% for $\sbrapp > 18.5$ and 64\% for $\sbrapp = 18.2$. This is ascribed to the explicit surface brightness cut in the Main Galaxy Sample to eliminate binary stars (\S \ref{sec: properties}). The overall fraction of galaxies affected by incompleteness at the faint end is small because the numbers observed drop well before the completeness drops.

At the bright end, objects will be missed either through being shredded into smaller objects by the deblender or, as mentioned, by being misclassified as an object other than a galaxy. Besides the marginal seeing example mentioned, in recent years evidence has also emerged for a class of galaxies, ultra-compact dwarfs \citep[UCDs, e.g.][]{phillipps:ucd}, which are affected by this. Again any effect on the overall LF for the apparent magnitude range investigated in this paper is unlikely to be significant because, as shown by \citet{blanton:lflowl}, few objects are observed in this regime.

The VAGC quality flag rejects objects which are bad deblends in the ranges $z < 0.01$ and $\mrlogh > -15$. Most bad deblends are likely to be fragments of large galaxies and would therefore be within this range, being both nearby due to the apparent size of the galaxy in which they are contained and intrinsically faint due to their small mass and therefore light compared to a galaxy. Objects without a spectrum will not be in the range but will also not be shown in the LF.

\citet{strauss:mainsample} show that in the range $-23 < \sbrapp < -24.5$, 35\% of the objects are bad deblends after applying their local versus global sky value cut. Therefore objects that are in this 35\% but not in the range rejected by the quality flag may remain in the LFs presented here. Accurately quantifying the number of objects for which this is the case would require detailed inspection of images in large numbers of LF bins, which is beyond the scope of this paper. Here we inspect the images for some of the outlying bins and plot the LF slices (but not the greyscale) only for bins containing more than 20 galaxies, for which any contamination is small. Bad deblends are therefore unlikely to cause any significant biases in the overall LFs. An exception is again made for $\tjpg$, where the cut is 5 galaxies per bin, because these were manually inspected for outliers by the JPG and so the shot noise becomes the limiting factor.

Objects missed altogether by the survey will mean that we are not measuring the true LF. However, studies \citep[e.g.][]{hayward:unimportance} have shown that low-surface brightness galaxies do not contribute significantly to the luminosity density of the universe, particularly in the range of brightness we study here. Similarly, UCDs have absolute magnitudes of $-13 \simlt M_B \simlt -11$ \citep{drinkwater:ucd} and so would not enter into our LFs.

\begin{figure*} \centering
\includegraphics[height=0.75\textwidth,angle=270]{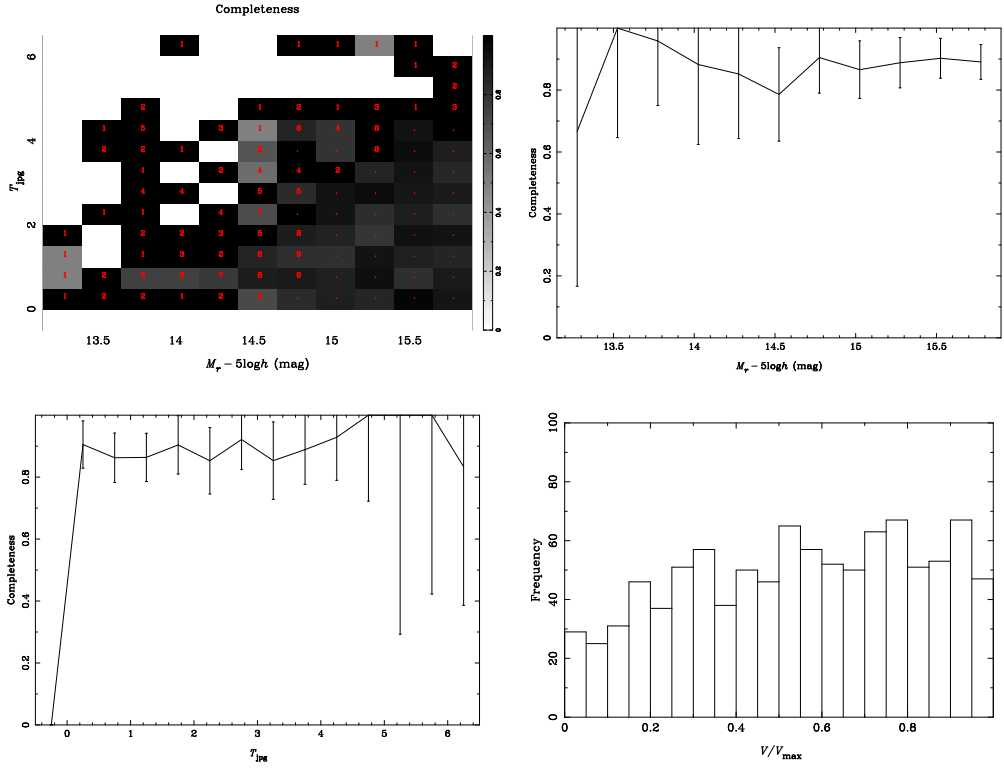}
\caption{Completeness values for the bivariate luminosity-$\tjpg$ distribution. The top left panel shows the function in a similar manner to the upper panels of Figs. \ref{fig: JPG bivlf}--\ref{fig: Density bivlf}. The top right and bottom left panels show the completeness as a function of apparent magnitude and $\tjpg$ respectively. The bottom right panel shows the $V/V_{\mrm{max}}$ statistic. The selection boundaries are the edges of the plot on both axes. \label{fig: JPG completeness}}
\end{figure*}

\begin{figure*} \centering
\includegraphics[height=0.75\textwidth,angle=270]{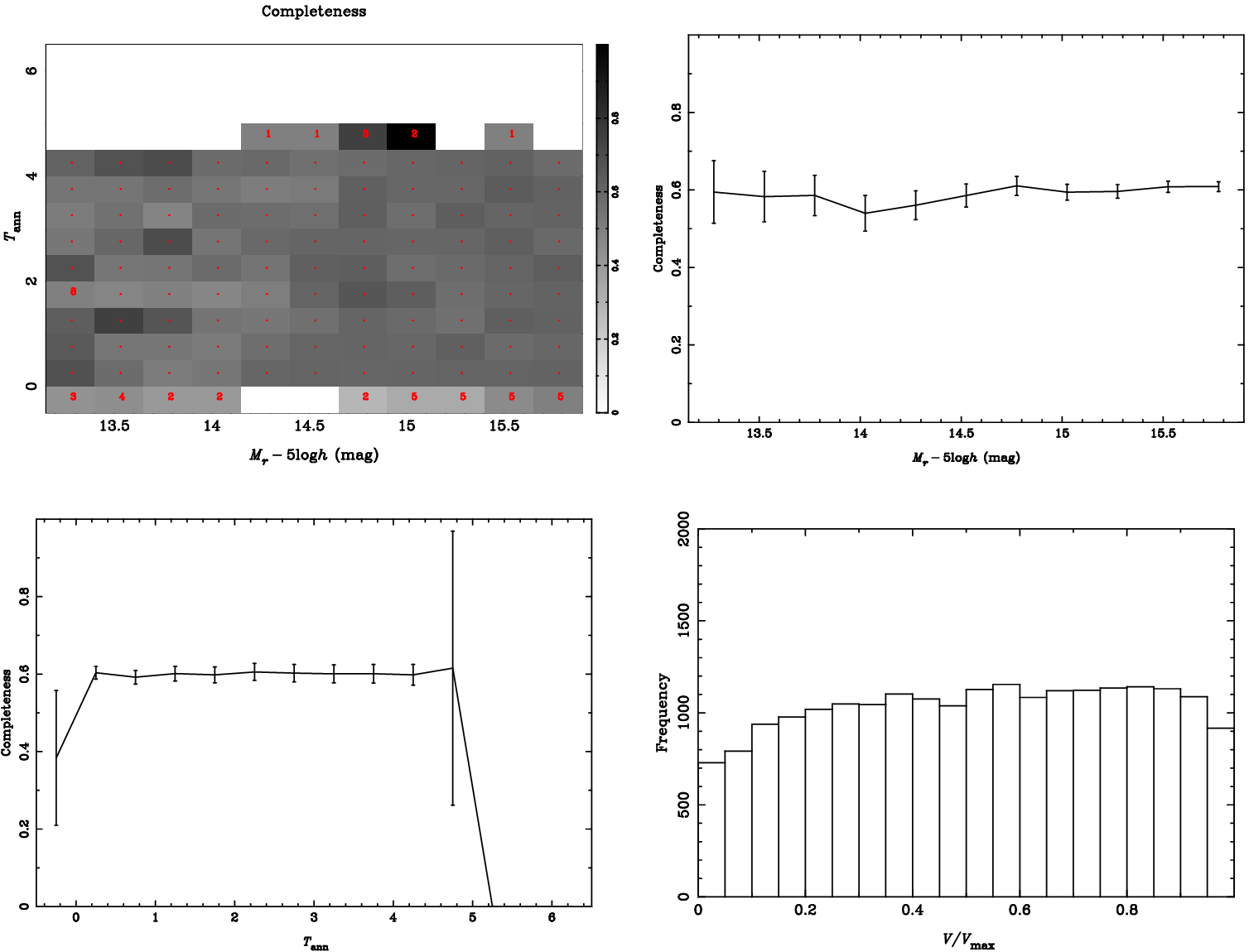}
\caption{As Fig. \ref{fig: JPG completeness} but for $\tann$. \label{fig: ANN completeness}}
\end{figure*}

\begin{figure*} \centering
\includegraphics[height=0.75\textwidth,angle=270]{figures/figure_A3.eps}
\caption{As Fig. \ref{fig: JPG completeness} but for $\ciinv$. \label{fig: invCI completeness}}
\end{figure*}

\begin{figure*} \centering
\includegraphics[height=0.75\textwidth,angle=270]{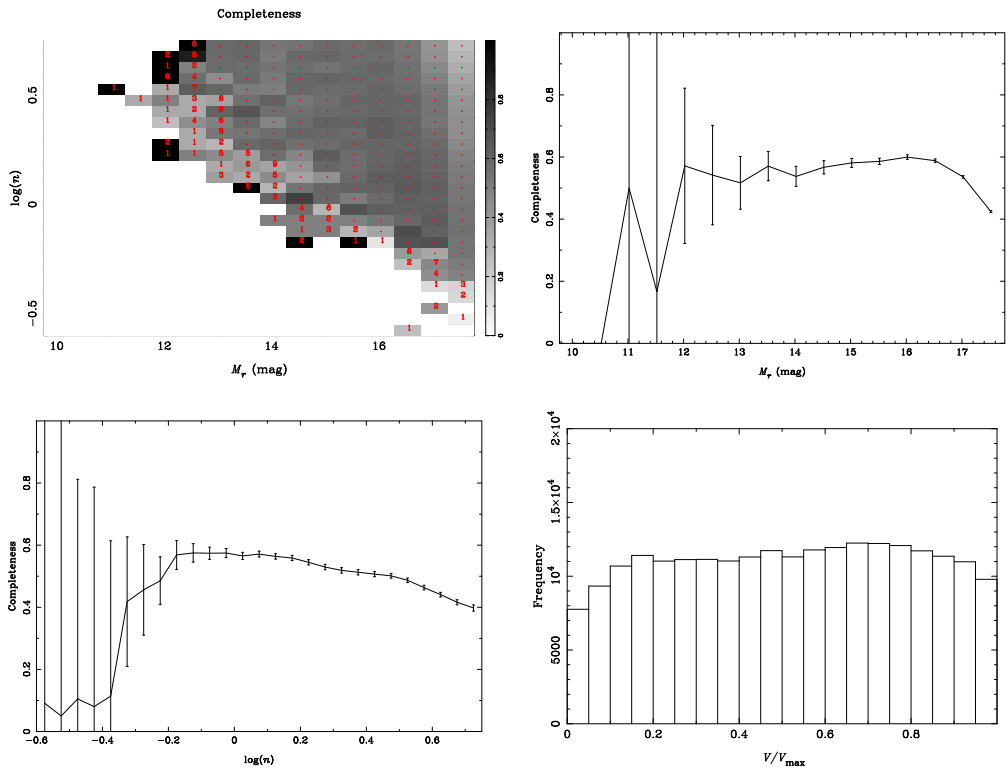}
\caption{As Fig. \ref{fig: JPG completeness} but for S\'ersic index $n$. \label{fig: Sersic completeness}}
\end{figure*}

\begin{figure*} \centering
\includegraphics[height=0.75\textwidth,angle=270]{figures/figure_A5.eps}
\caption{As Fig. \ref{fig: JPG completeness} but for $\sbr$. \label{fig: SB completeness}}
\end{figure*}

\label{lastpage}

\end{document}